\documentclass[usenatbib]{mn2e}
\usepackage{color} 
\usepackage{graphics}
\usepackage{epsfig}
\usepackage{natbib}

\begin{document}

\title[Edge-on galaxies in MaNGA] 
{An integral field spectroscopic study of stellar and ionized gas properties 
around  edge-on disk galaxies in the stellar mass range $9<\log M_* < 11$ }    

\author [G.Kauffmann] {Guinevere Kauffmann$^1$\thanks{E-mail: gamk@mpa-garching.mpg.de},
Richard D'Souza$^2$, Antonela Monachesi$^3$\\
$^1$Max-Planck Institut f\"{u}r Astrophysik, 85748 Garching, Germany\\
$^2$Vatican Observatory, Specola Vaticana, V-00120, Vatican City State\\
$^3$Universidad de La Serena, Avda. Raul Bitran 1305, La Serena, Chile}

\maketitle

%===================================
\begin{abstract} 
We analyze  the stellar light, 4000 \AA\ break and emission line
profiles of 82 edge-on disk galaxies from the MaNGA  survey.
We characterize the stellar light profiles perpendicular to the disk
plane using two parameters: a) the power law  slope of the  thick disk
component, 2) the transition radius where the profile flattens.  The 4000
\AA\ break profiles perpendicular to the  plane are characterized by
the number of significant changes in slope (breaks)  and
by the change in D$_n$(4000) from inner to outer disk.
The slope  correlates tightly  with the
stellar mass of the galaxy over the stellar mass range $10^9< \log M_*
< 10^{10} M_{\odot}$. More massive galaxies have more
extended thick disks. The slope and transition radius   exhibit
large scatter for galaxies more massive than $10^{10} M_{\odot}$.  Half the sample have older stellar
populations in their thick disks, a third have flat D$_n$(4000) profiles
and  15\%  have younger thick disks.  The D$_n$(4000) profiles exhibit
as many as 4 separate breaks. There are more breaks  in massive galaxies
with bulges and more extended thick disks. This may indicate that the breaks are produced
by more frequent accretion events in such systems. 
The extraplanar H$\alpha$ EQW  correlates most strongly with the specific star
formation rate of the galaxy, and the [OII]/H$\alpha$ ratio  increases
with distance from the disk plane. This increase is most apparent for
massive galaxies with extended thick disk components and low  SFR/$M_*$.
These findings support the hypothesis that the larger [OII]/H$\alpha$
ratios may be caused by ionization from evolved stars. 
\end {abstract}

\section{Introduction}

The study of edge-on disk-dominated galaxies provides important information
about the 3-dimensional distribution of stars, gas and dark matter in these
systems. Early work in the 1980's (e.g. Van der Kruit \& Searle 1981, Van
der Kruit 1988) was based on surface photometry from photographic plates
that reached limiting surface brightnesses of 22 magnitudes arcsec$^{-2}$
in the optical.  To model the light profiles, these authors invoked an
exponential distribution of stars in the radial direction and adopted an
isothermal sheet approximation to describe the  structure of the disk in the
vertical direction. The radial structure was proposed to be regulated by the
angular momentum structure in the proto-galaxy (Freeman 1974; Gunn 1982),
while the vertical thickness of the disk was regulated by so-called secular
processes, including heating by molecular clouds and spiral arms, and was
believed to carry no information about the structure of the proto-galaxy.

The thin disk scale height of nearby disk galaxies is typically less
than 1 kpc.  The existence of vertically extended  disk stellar components
emerged into the mainstream from studies of  counts of stars as a function of
distance from the plane of the Milky Way (Gilmore \& Reid 1983). 
In addition, photometric studies had by then  revealed that some  nearby edge-on galaxies had  
thickened disks  (Van der Kruit \& Searle 1981). Later, 
deep imaging of edge-on  early-type spiral (Wu et al 2002) and lenticular
galaxies (Pohlen et al 2004) showed  that the  thick disk components were distinct 
from the thin disk.

Dalcanton \& Bernstein (2000) carried out  deep
imaging studies reaching limiting magnitudes of 29 mag arcsec$^{-2}$ at
opical wavelengths.  These authors selected  49 extremely late-type, edge-on
disk galaxies from the Flat Galaxy Catalog of Karachentsev et
al. (1993). The sample was chosen to have no apparent bulges or optical
warps so that the galaxies represent undisturbed ``pure disk'' systems. The
main result of this study was that such thin disks are embedded in a low
surface brightness red envelope extending to at least  5 vertical disk scale
heights above the galaxy mid-plane, with a radial scale length that appeared
to be uncorrelated with that of the embedded thin disk. The color of the
red envelope was found to be similar from galaxy to galaxy, even when the
thin disk is extremely blue, and was consistent with a relatively old ($>$6
Gyr) stellar population that was not particularly metal-poor (Dalcanton \&
Bernstein 2002).  These authors argued that the ubiquity of the red stellar
envelopes implied that the formation of the thick disk is a nearly universal
feature of disk formation and is not necessarily connected to the formation
of a bulge. One caveat of this work was that the available K-band photometry
was relatively shallow and did not allow for very precise constraints on
stellar age or metallicity.

There have also been attempts to study how the importance of the  thick
disk component depends on the mass or circular velocity  of the galaxy,
Yoachim \& Dalcanton (2006) studied the scale heights of the thin and thick
disk components of late type galaxies as a function of circular velocity.
The average  scale height of the thin disk increased from $\sim$ 300 pc
for galaxies with $V_c \sim 50$ km/s to 1 kpc for galaxies with $V_c \sim
200$ km/s, while the scale height of the thick disk increased from 1 to 3
kpc over the same circular velocity range. The ratios of the thick and thin
disk scale heights and stellar masses  were found to decrease with circular
velocity. It should be noted that Comeron et al (2011, 2014) repeated this
work using kinematic information to deconvolve the ``hot'' and the ``cold''
components of the disk. If the mass of the kinematically hot central component
of the galaxy was added to the thick disk, the trend in thin-to-thick disk
mass ratio as a function of $V_c$  was nearly absent.

Spectroscopic studies of the vertical structure of edge-on disk galaxies,
either using slits oriented along the minor axis of the disks, or using
integral field units (IFU) have generally focused on the properties of
the ionized extra-planar gas detected in emission. Lehnert \& Heckman (1996)
carried out a study of 50 IR-selected edge-on starburst galaxies and found
that extra-planar emission line gas is a common features of these systems.
The filamentary and shell-like morphology of the gas, the increase in line
width along the minor axis of the galaxy, and line ratios indicative of shock
ionization pointed to a galactic wind origin for this gas.  The strength
of the emission lines was also found to correlate with the IR luminosity of
the galaxies.

More recently, integral field unit (IFU) spectroscopic
surveys of galaxies in the local Universe have allowed the
extra-planar gas to be studied in large samples of normal disk galaxies. 
Ho et al (2016) studied a sample of 40 local main-sequence edge-on disk galaxies
observed as part of the SAMI Galaxy Survey, and found that emission line
kinematics characteristic of galactic winds were more common in galaxies
with high star formation surface densities and in galaxies with spectroscopic
signatures of recent starbursts. Jones et al (2016) derived average emission
line luminosity profiles from stacked spectra of galaxies from the MaNGA IFU
survey and found the largest differences in emission line profiles
and emission line ratios  occurred when galaxies were split by stellar mass.
In particular, the higher oxygen line to Balmer emission line ratios
in the outskirts of massive galaxies were taken as evidence for a change in
the ionization source of the lines from massive young stars in HII regions
in the inner disk, to ionizing radiation from evolved stars 
in the outer regions.

Theoretical modelling of the origin of thick disk and halo components of galaxies
has advanced dramatically in recent years.  Early studies examined specific 
disk heating mechanisms such as heating by dissipationless mergers of satellites  (Toth \& Ostriker 1992,
Quinn, Hernquist \& Fullagar 1993, Velazquez et al 1999) or dissipative accretion of gas-rich stellites orbiting 
in the same plane as the main disk (Bekki \& Chiba 2001, Brook et al 2004). This was followed by
an era where thick disk and halo formation processes were examined in hydrodynamical
zoom simulations of individual galaxies, where the accretion processes follow the expectations of 
dark matter halo assembly  in a $\Lambda$CDM cosmolgy (Brook et al 2012, Abadi et al 2013)

Most recently, it has been noted that thick disk formation processes can differ substantially
depending on the physical treatment of star formation and supernova/AGN feedback processes.
For example, Yu et al (2021) have studied thick stellar disc formation in Milky
Way-mass galaxies using 12 FIRE-2 cosmological zoom-in simulations and show how
that most of the mass is formed during periods of bursty star formation when
supernova-driven outflows drive a large amount of gas out to large distances
above the galactic plane. Pinna et al (2024) analyzed thick disk assembly
in the AURIGA simulations and found that there is early in situ formation,
but there is  later growth driven by the combination of direct accretion of
stars from satellite infall, some in situ star formation fueled by mergers,
and dynamical heating of stars. Elias et al (2018) have  also noted that there is  substantial
scatter from one halo to another within one simulation set and this has been verified
observationally for Milky Way mass galaxies (Merritt et al 2016). This means
large samples of galaxies are crucial in  making  fair comparisons between
simulations and data.

We note that there  have been almost no studies that have used IFU data to carry out a
combined analysis of the stellar and ionized gas properties of the same
galaxies. 
In this paper, we analyze a sample of 82 edge-on disk galaxies drawn from
the Mapping Nearby Galaxies at APO (MaNGA surevy). We stack the spectra
as function of vertical scale height above/below the disk and we are able
to trace the stellar light distribution out to typical distances of 10
kpc. As we will show, this is usually far enough to measure the radius at
which the stellar light profiles start to flatten because of the increasing
contribution of light from the  stellar halo (Gilhuly et al 2022). We are
also able to measure 4000 \AA\ break profiles out to distances of 4-6 kpc
above and below the disk as a probe of stellar population age,
and  strong emission lines
[OII]$\lambda$3727 and  H$\alpha$ as a probe of the ionized gas. The goal
is to study the corelations between the properties of the extra-planar
stars and gas with  the  properties of the main galaxy, such as mass,
bulge-to-disk ratio and specific star formation rate, as well as to 
understand how the extra-planar components correlate with each other.
Although this sample is not extremely large, we will show that it is sufficient
to examine trends in the two-dimensional space of  galaxy stellar mass and star formation
rate.

Our paper is organized as follows. In section 2, we describe the selection
of the galaxy samples used in the analysis. In section 3, we describe 
the methodology used for creating stacked spectra and deriving profiles
for quantities pertaining to the stellar light and to extra-planar ionized
gas. Results are presented in Section 4 and are summarized and discussed
in Section 5.

\begin{figure*}
\includegraphics[width=210mm]{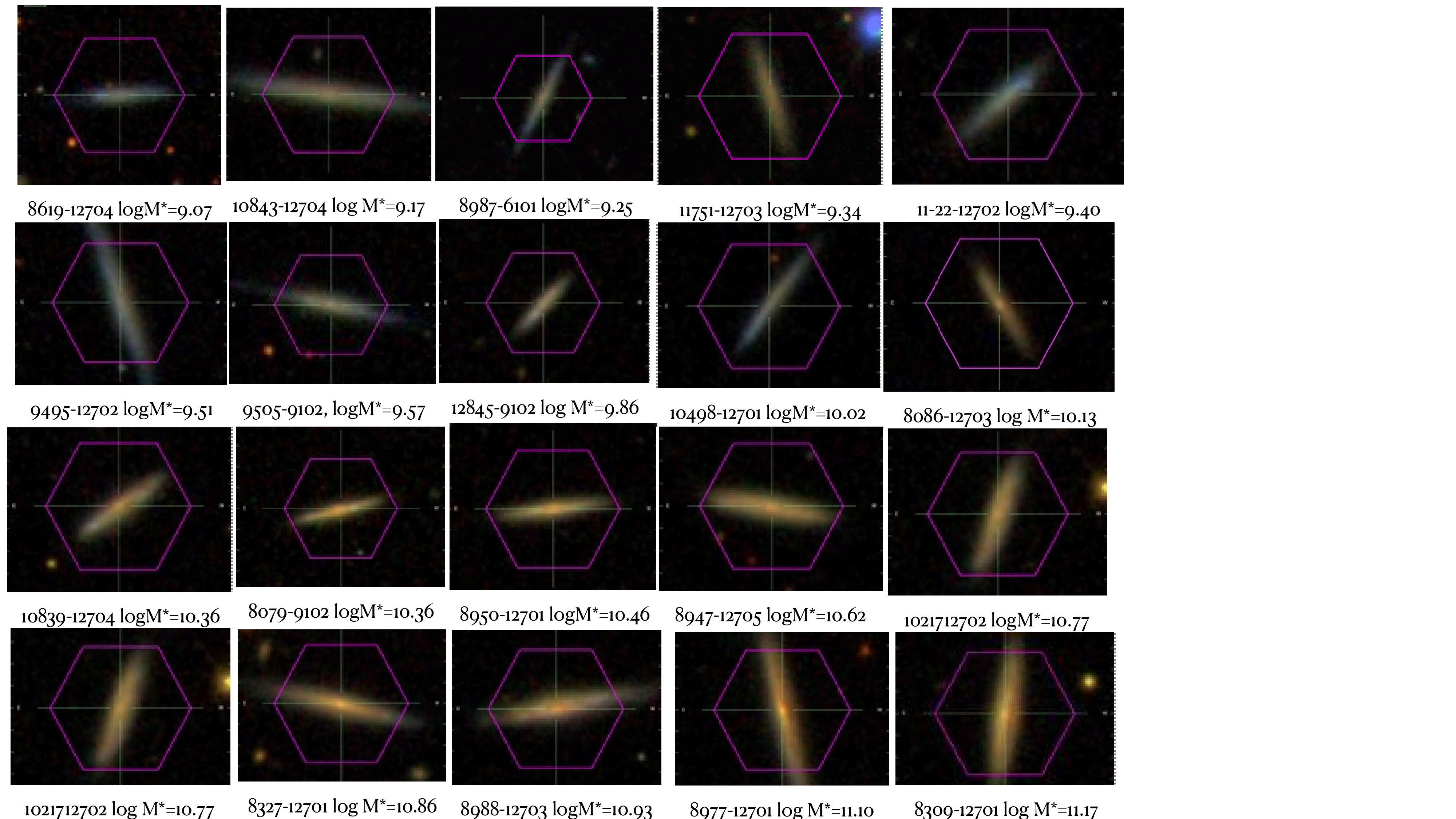}
\caption{A gallery of $g,r,i$-band cutout images of edge-on galaxies from 
the sample with Sersic index less than 2.5 is shown. The galaxies are arranged
in order of increasing stellar mass. Each image is labelled with 
its  MaNGA-id numbers and its stellar mass.
\label{models}}
\end{figure*}

\begin{figure}
\includegraphics[width=90mm]{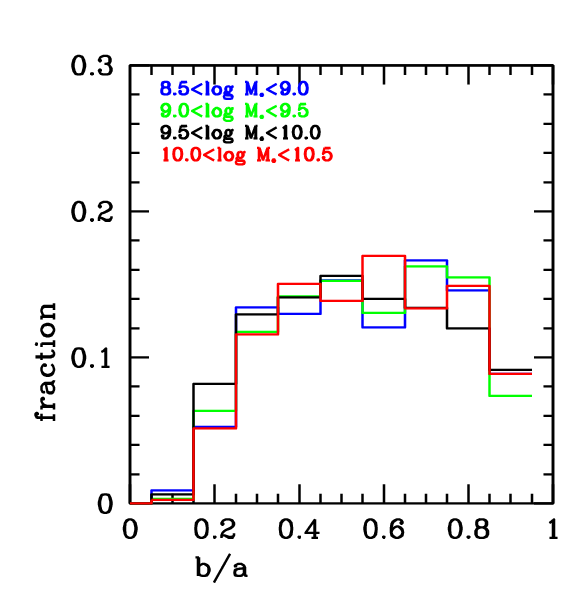}
\caption{The distribution of
$b/a$ in bins of stellar mass in the parent sample.
\label{models}}
\end{figure}

\section {The galaxy sample}

We use the final version of the MaNGA data (Abdurro'uf et al. 2022). MaNGA
was a survey carried out as part of the  SDSS-IV survey (Sloan Digital Sky
Survey) using IFU observations of galaxies to produce spatially resolved
spectroscopic data (Gunn et al. 2006; Bundy et al. 2015; Drory et al. 2015;
Blanton et al. 2017; Aguado et al. 2019; Abdurro'uf et al. 2022). MaNGA
observed approximately 10,000 galaxies, selected as an unbiased sample in
terms of stellar mass over the range $M_*>10^9 M_{\odot}$ and environment
(Law et al. 2015; Yan et al. 2016; Wake et al. 2017).  The primary sample
in MaNGA consists of  of about 5000 galaxies  selected with IFU coverage
out to $1.5 R_{eff}$  . The secondary sample of the main subgroups includes
about 3300 galaxies, observed by the IFU bundle out to 2.5 $R_{eff}$.

The IFU bundles of the MaNGA survey consist of 19-127 fibres with hexagonal
shape covering 12-32 arcseconds in diameter on sky. The spatial resolution of
MaNGA is 2.5 arcsec , which corresponds to 1.3-4.5 kpc for the primary sample
and 2.2-5.1 kpc  for the secondary sample. The BOSS spectrographs used by
MaNGA provide spectra from 3600 to 10300 \AA\  with a spectral resolution 
$R \sim 2000$ (Smee et al. 2013; Yan et al. 2016). The observations
were conducted to reach a signal-to-noise ratio (S/N) in the stellar continuum
of 14-35  per spatial sample, which required about 3 hours net integration
for each target.

In this paper, we focus on a sample of 82 edge-on disk galaxies selected from
the MaNGA sample.  As discussed in Beom et al (2022), selection of suitable
samples of  edge-on galaxies requires two main steps. First, stellar
masses and structural parameters such as position angles, ellipticities,
and half-light radii for all galaxies in the MaNGA sample are available
from the MaNGA drpall file; in all cases, the structural parameters are
obtained from the SDSS elliptical Petrosian apertures.  In the initial cut,
all galaxies with major-to-minor axis ratio b/a less than 0.35, Sersic indices
less than 2.5, redshifts greater than 0.01 and stellar masses in the range
$10^8<M_*<10^{11} M_{\odot}$ are selected. This results in a sample of 481
galaxies. The second step involves visual inspection of the images to find
those objects where the IFU covers most of the disk, and  where there are no bright
stars or galaxies in the vicinity where scattered light could contaminate the
off-axis measurements of flux from the galaxy. In addition, we eliminate merging
galaxies and galaxies in dense environments such as groups and clusters where
it would be difficult to disentangle the low surface brightness emission from
different systems. These cuts are fairly stringent and result in a sample
of 68 galaxies. A gallery of $g,r,i$ cut-out images is shown in Figure 1.
The galaxies are arranged in order of increasing stellar mass and the purple
hexagons show the spatial  coverage of the MaNGA IFU.

The sample cut on Sersic index and $b/a$ is designed to select only
edge-on, disk-dominated galaxies for study. In Figure 2, we plot the distribution of  
$b/a$ in bins of stellar mass and find that these are almost identical,   
indicating that our cut is selecting a morphologically uniform population
of objects. In order to quantify the effect of a more prominent bulge
component on the properties of the thick disk and the inner halo, we eliminate 
the cut on Sersic index and visually examine all  galaxies with $b/a < 0.40$ 
and Sersic index greater than 2.5. We find that most systems are irregular/merging
objects or galaxies with a central bulge where the disk is not edge-on. 
We are able to  extract a sample of 14 relatively ``clean'' edge-on  galaxies 
with larger central light concentration for comparison with the main sample
of 68 galaxies. A gallery of images from this sample arranged by stellar
mass is shown in Figure 3. Our final sample thus consists of a total
of 82 edge-on galaxies.  

\begin{figure*}
\includegraphics[width=135mm]{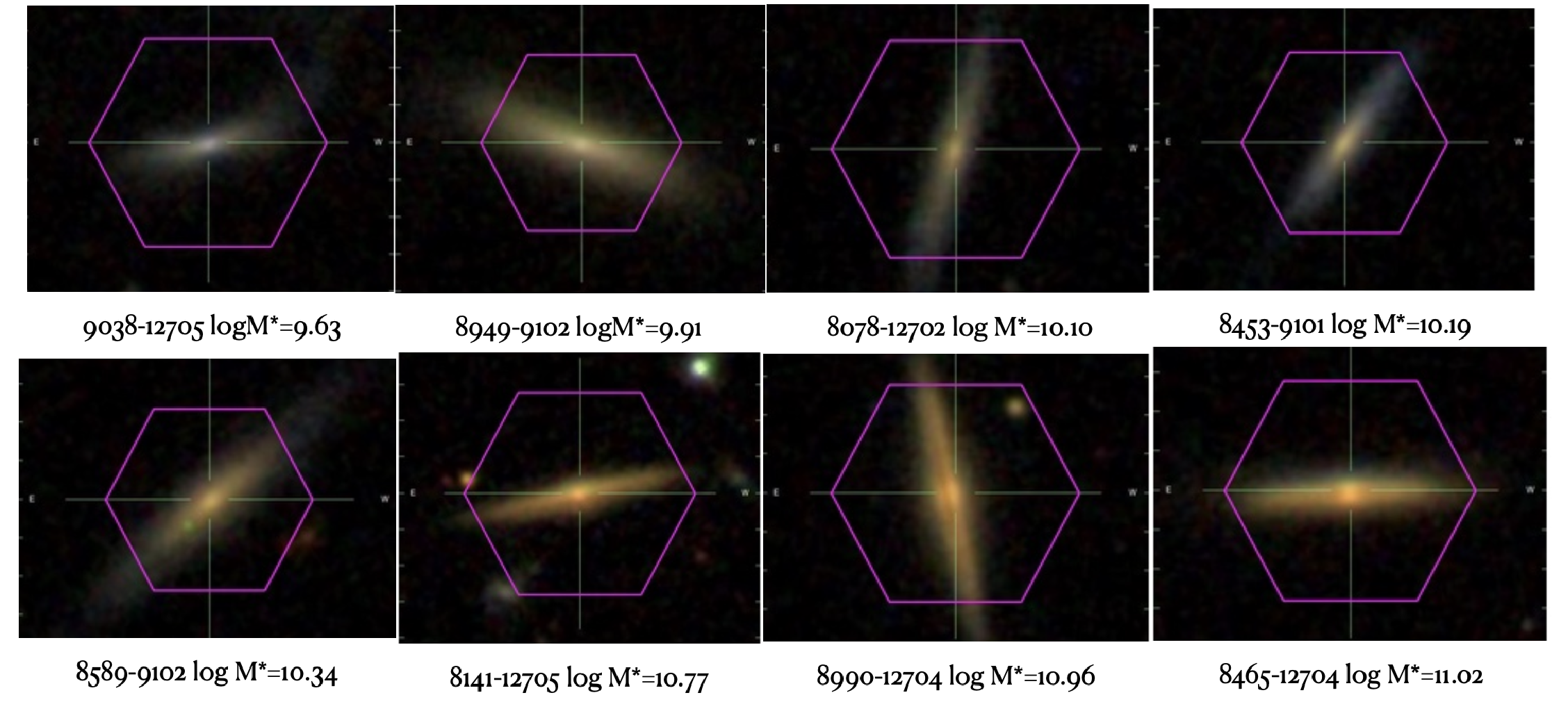}
\caption{Same as Figure 1, except for galaxies with 
Sersic index greater  than 2.5.
\label{models}}
\end{figure*}

\section {Analysis methods}

The primary MaNGA data products are composed of 3-D calibrated data cubes 
produced by the data reduction pipeline (DRP). In this section, we discuss how we combine  
spectra from the MaNGA cube file to calculate the stellar light and D$_n$(4000) break  
profiles of the galaxies in our sample , as well as average emission line strengths
in the thick disk and inner halo.

We work with the SPX-MILESHC-MASTARSSP LOGCUBE and MAPS files, which contain spectra and analysis
products for each individual spaxel with $S/N>1$. the orientation of the edge-on thin disk                     
is determined from the g-band weighted mean flux (SPX-MFLUX) measurements for each spaxel. 
We perform  successive least squares linear fits to subsets of the spaxels, starting from 
the 10\% brightest central spaxels, and then including fainter and fainter spaxels iteratively. 
\footnote{
The central spaxel is always defined using the MaNGA coordinate system as given in
the Data Analysis Pipeline.}
During each fitting procedure, we calculate the position angle of enclosed light
using the linear fit, as well as maximum radial extent from the central spaxel.

The reason the iterative procedure is needed, is because many galaxies have inner
light concentrations that are misaligned with the more extended region of the disk.
As fainter and fainter spaxels are included, the fit will start to orientate with
the low surface brightness light away from the disk and  the maximum
radial extent of the spaxels will no longer increase significantly. The iterative fitting
procedure almost always converges to a stable estimate of the orientation of the disk in a 
region where the maximum radial extent is still increasing on each iteration, but
the estimated disk position angle remains fixed.  

The next step is to create stacked spectra as a function of perpendicular distance from
the disk. We create median stacked spectra in order to reduce the influence of outliers
and we estimate the error on the median using a standard bootstrap error estimation 
method with 100 replications. Our stellar continuum flux estimation is carried out over
the wavelength interval from 5015 to 6290 \AA. This wavelength range is chosen because it 
does not contain any strong emission lines and it is also in the region of the spectrum
with highest $S/N$. The spectra are arranged in order of increasing perpendicular
distance from the disk and the stack is first carried out in bins of 20 spectra
at a time. If the  S/N in the continuum flux is less than  70, more spectra are added to
the stack until $S/N=70$ is reached. The stacking process is stopped once  there are no 
more spectra left to stack, or the $S/N$ of the stack falls below 10.

In Figure 4, we show the profiles for the same disk-dominated  galaxies pictured in Figure 1.  
The logarithm of the average flux per spaxel is plotted as a function of distance from
the disk plane in units of kiloparsec. The cyan shaded region shows the error estimate
on the flux. As can be seen, the outer disk profiles of low mass galaxies (top and second rows)
 are well-characterized
by a steeply declining power-law. In some cases, the profile appears to break or flatten.
As the stellar mass of the galaxy increases, the outer disk profiles have a shallower slope
and a  flattening at large radius is clearly found in almost 
all galaxies (third and bottom rows). We will
describe how we place the profile trends  on quantitative footing through profile fitting 
in the next section. 
\footnote{The profiles of the sub-sample with Sersic index greater than 2.5 are
quite similar in terms of basic trends such as slope of the thick disk profile
as function of stellar mass and the flattening in the outer regions, so we do not
show a separate plot for these objects.}  

\begin{figure*}
\includegraphics[width=160mm]{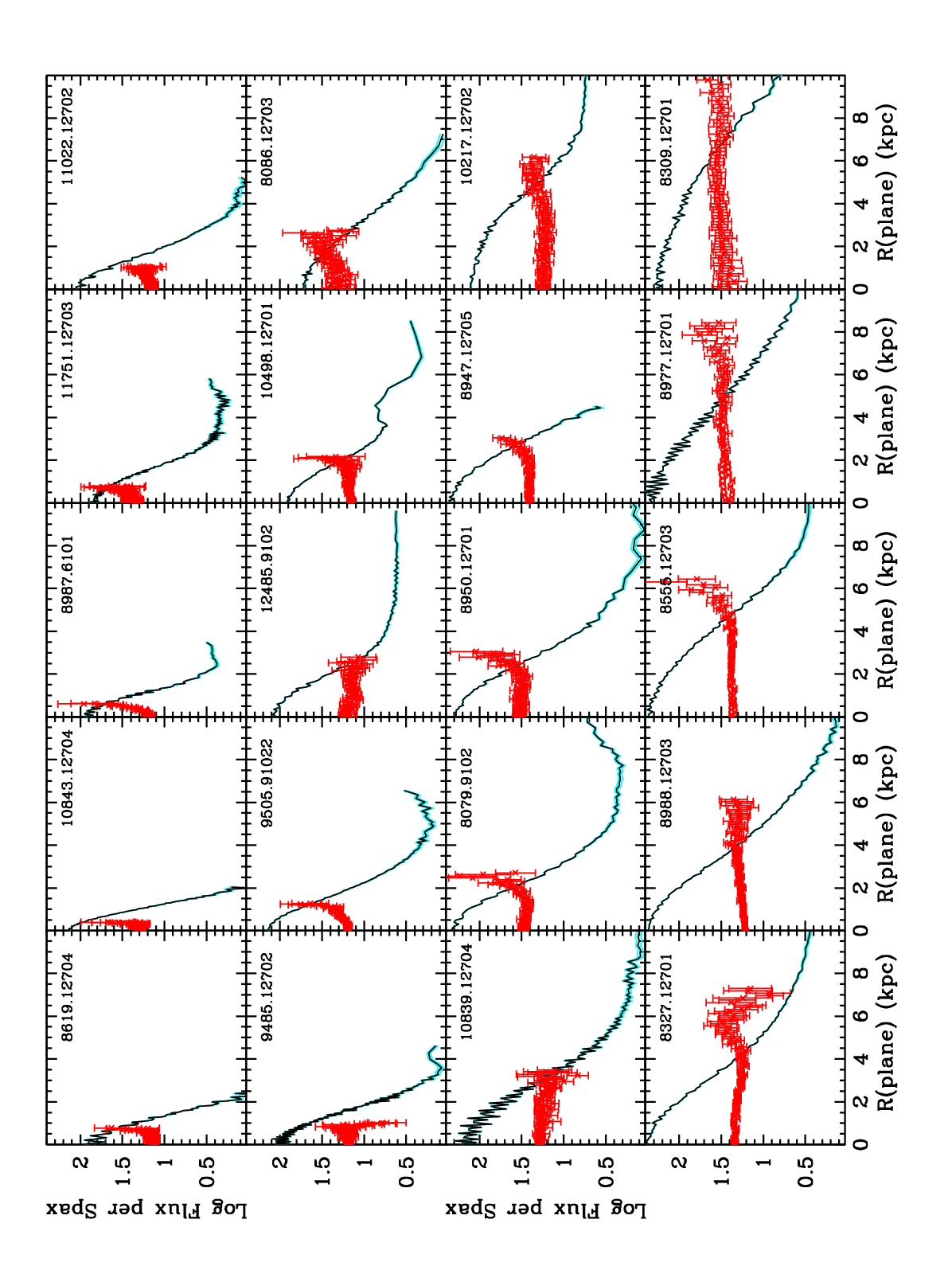}
\caption{ Surface brightness profiles for the same disk-dominated  galaxies pictured in Figure 1,
arranged in order in increasing stellar mass, from left to right and from top to bottom. 
The logarithm of the average flux per spaxel is plotted as a function of distance from
the disk plane in units of kiloparsec. The cyan shaded region shows the error estimate
on the flux.  D$_n$(4000) profiles are  plotted
as red points with error bars over the  range where the error remains smaller than 0.2.
Note that the flux units have been scaled so that we are able to show the 
surface brightness and D$_n$(4000) profiles using the same values on the y-axis of
each panel in  the figure.
\label{models}}
\end{figure*}

We also calculate the 4000 \AA\ break index D$_n$(4000) and plot
these as red points with error bars over the  range where the error remains smaller than 0.2.
D$_n$(4000) in normal galaxies varies over the range between  1.1 for galaxies with very
young stellar populations, where a recent burst of star formation has occurred, to 
values of 2 for galaxies with old, metal-rich stellar populations (see Kauffmann et al 2003;
Kauffmann 2014). As can be seen, our  D$_n$(4000) measurements always fall within this range.  
The majority of galaxies have D$_n$(4000) profiles that are flat in the inner disk and then 
increase in the power-law  thick disk region, in agreement with the findings of 
Dalcanton \& Bernstein (2000). In most galaxies, we are not able to obtain a high S/N
measurement of the 4000 \AA\ break in the region where the surface brightness profiles
flatten. In a few galaxies, the D$_n$(4000) profiles show a break towards smaller
values at large distances and there also may be more than one break in the profile. We
will also quantify the D$_n$(4000) trends in more detail in the next section.

Finally, in order to detect the emission lines in the stacked spectra of the thick disk and
halo-transition region at high enough S/N, we stack spectra in  3 regions: 1) the
inner disk region where the surface brightness profiles are flat, 2) the thick disk
region where the surface brightness profiles are declining with a fixed power law 
slope, 3) the halo transition region where the surface brightness profiles are flattening. 
An example of how these stacks are made is shown for the galaxy 10839-12704
in Figure 5. The cyan, red and yellow coloured regions show the regions of the galaxy
within 80\%, 60-80\% and 20-60\% of the peak surface brightness. The green, black and red points
indicate the positions of the spaxels used to create the stacked spectra in the
inner disk, the thick disk and the halo. The resulting spectra  are plotted  in 4 different wavelength
ranges containing the strongest emission lines in Figure 6. The top row of the
figure shows the spectral stacks over the wavelength range covering the H$\alpha$ and
[NII] lines as a well as the [SII] doublet. The second row shows the region covering
the high-ionization [OIII] line. The third and fourth rows cover the lower ionization
[OII] and [OI] lines and the bottom row covers the 4000 \AA\ break region of the spectrum.  

\begin{figure}
\includegraphics[width=90mm]{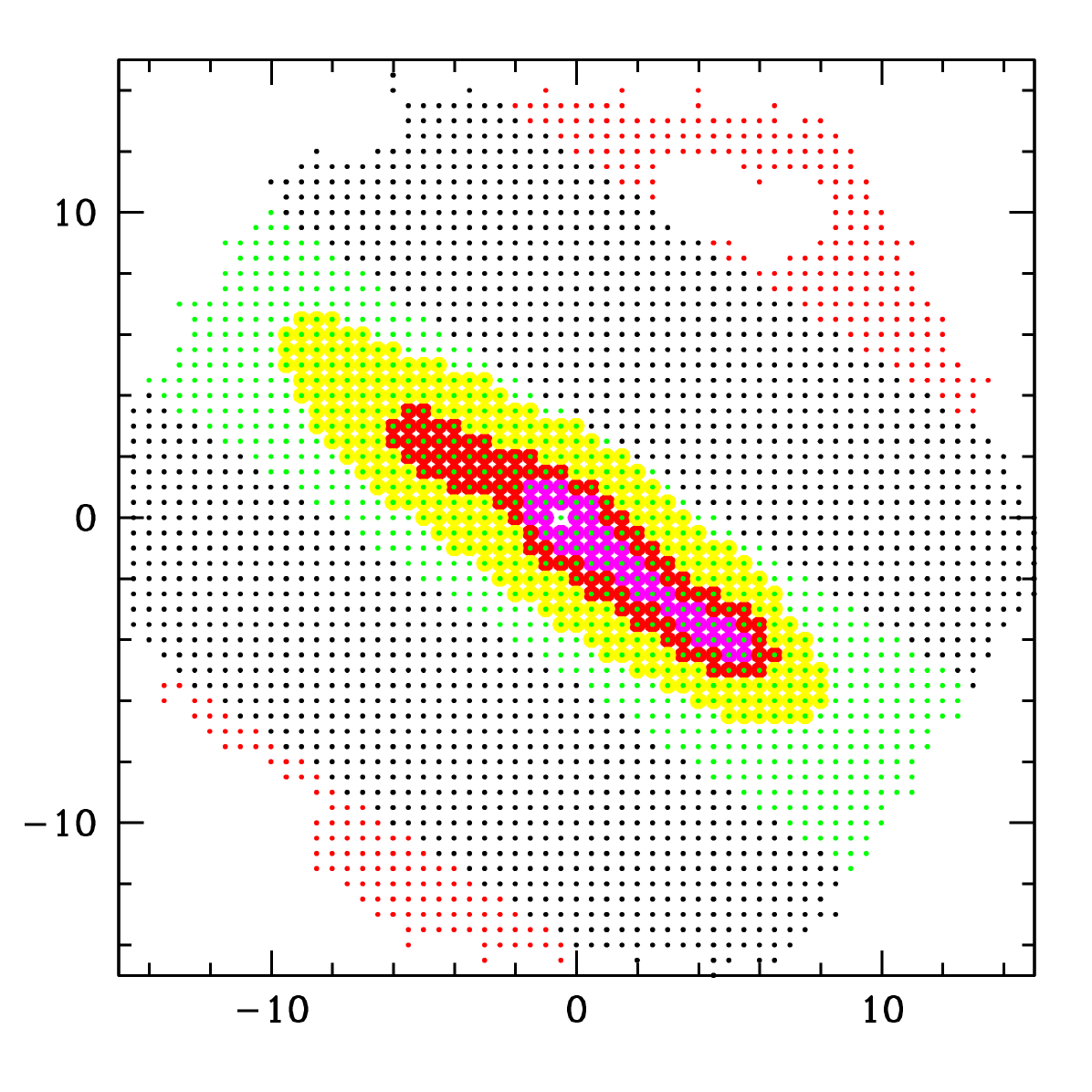}
\caption{
An example of how  stacks in the inner disk, the thick disk and the halo are made is shown for the galaxy 10839-12704.
 The cyan, red and yellow coloured regions show the regions of the galaxy
within 80\%, 60-80\% and 20-60\% of the peak surface brightness. The green, black and red points
indicate the positions of the spaxels used to create the stacked spectra in the
inner disk, the thick disk and the halo, respectively.
\label{models}}
\end{figure}

\begin{figure*}
\includegraphics[width=160mm]{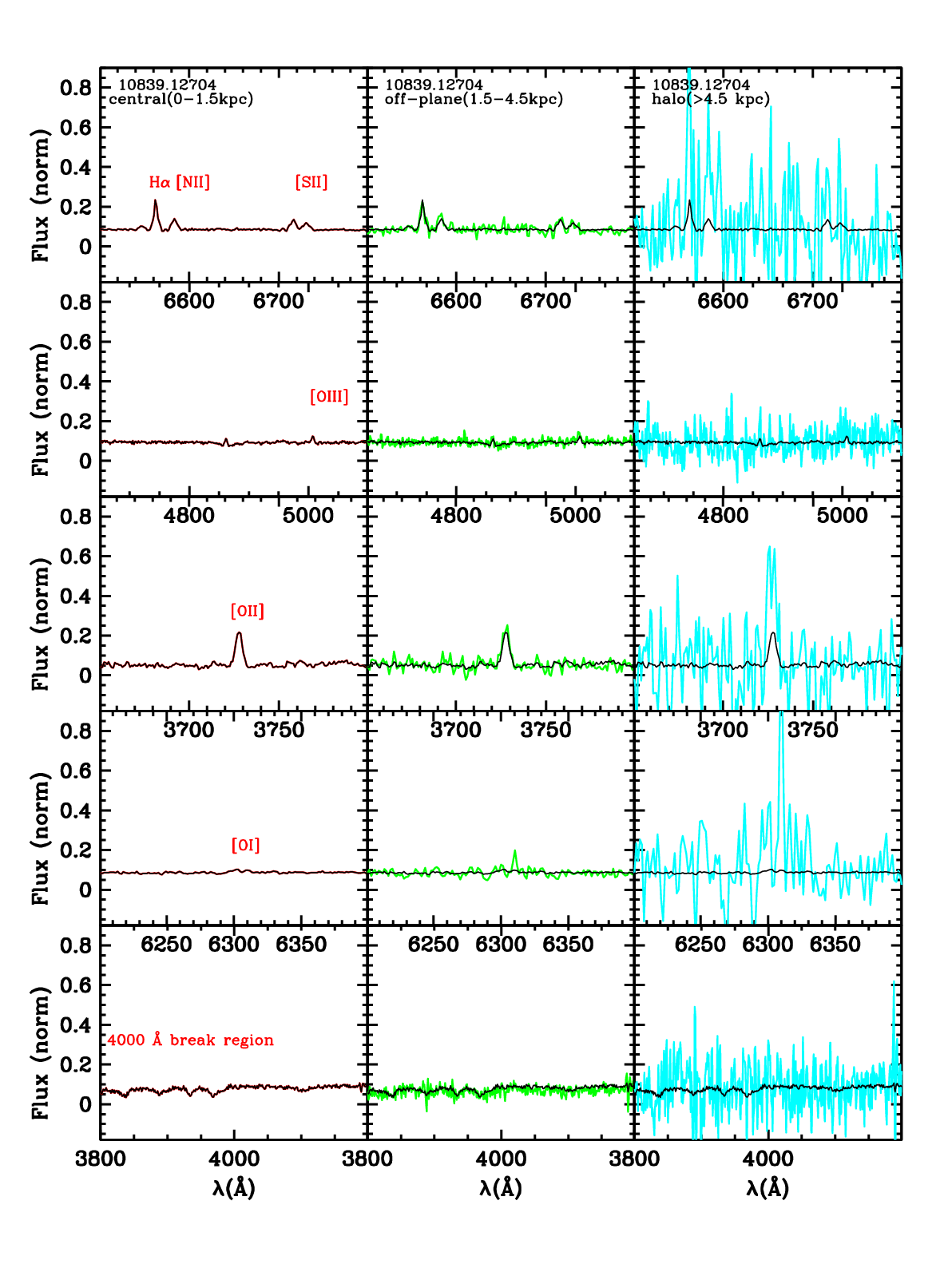}
\caption{Stacked spectra for the galaxy 10839-12704 in 4 different wavelength
ranges containing the strongest emission. The top row of the
figure shows the spectral stacks over the wavelength range covering the H$\alpha$ and
[NII] lines as a well as the [SII] doublet. The second row shows the region covering
the high-ionization [OIII] line. The third and fourth rows cover the lower ionization
[OII] and [OI] lines and the bottom row covers the 4000 \AA\ break region of the spectrum.
The columns show  spectra in  3 regions: 1) the
inner disk region where the surface brightness profiles are flat, 2) the thick disk
region where the surface brightness profiles are declining with a fixed power law
slope, 3) the halo transition region where the surface brightness profiles are flattening.
The black spectrum that is shown in all panels is the inner disk spectrum, while the
red green and cyan spectra are for inner dis, thick disk and halo. 
\label{models}}
\end{figure*}

The stacked spectra plotted in the left columns are for the inner region of the
disk seen edge-on, the middle column shows the stacked spectra in the thick disk region
and the right column shows the spectra in the transition regions between thick disk and halo.
The [OII] and H$\alpha$ emission lines are almost always detected in the thick disk
region and also with reasonable frequency in the inner halo. The [OIII] line is almost
never detected outside the inner disk. The [OI] line is detected in a small subset of
galaxies out to large radius from the disk. In this paper, we will focus on the 
[OII] and H$\alpha$ lines as tracers of the ionized gas.

Because the  spectra in the outer disk regions are very noisy and because the line shapes are
quite often complex, we use a non-parametric method to calculate the emission line equivalent
widths. We define two continuum bands of width 20\AA\ located  symmetrically  on either side of the emission line. 
The adopted continuum bands are located further away from the rest-frame wavelength of
the line in the outer disk and in the inner halo stacks, because the emission lines are
frequently broader in these regions. We calculate $F_{cl}$ and $F_{cr}$, the continuum flux
on the left  and the right side of the lines, along with errors estimated by bootstrap resampling
of the spectral measurements. Within the  region that is assumed to contain the line emission, we
locate the maximum flux point, and then integrate the flux to the left and to the right of this point
until it drops below the continuum level by 2$\sigma$ for the first time. This procedure is
repeated 100 times on bootstrap resampled spectra to calculate a median EQW as well as the
16-84th percentile confidence ranges. Our results will be presented in the next section.

\section {Results}

\subsection{Parameters of the stellar light profiles and their scalings}

Many past papers on the structure of  edge-on disk galaxies have focused on
accurate methods for deconvolving the thin and thick disk components. The IFU
observations of edge-on galaxies studied here have limited
($\sim$ 1 kpc) spatial resolution
and we will not adopt this approach. Rather than attempting a 3-zone
decomposition into thin, thick disk and halo components,  only a two-zone
definition of power law for the thick disk at distances greater than
1-1.5 kpc  above the plane
and flattening for the stellar
halo is made.   

 As discussed in the previous section,
the average surface brightness profiles as a function of perpendicular distance
to the disk plane are well-characterized by a power-law region that extends
from $\sim$ 1 kpc to  typical distances of  3-7 kpc, before flattening.
We will thus paramertize these profiles with two quantities: 1) The slope $b$
in the power-law part of the profile, 2) R$_{break}$, the break radius. where
the profile lies more than $3 \sigma$ above the best-fit power law for four
successive radial bins.

In Figure 7, we show how these two profile parameters depend on the stellar
mass and the logarithm of the  specific star formation rate $\log SFR/M_*$
for 82 galaxies in our sample. The disk-dominated galaxies are plotted as
black circles , while the sample with higher Sersic indices are plotted as
red triangles. In these plots, we have used stellar masses and star formation
rates from the MaNGA Pipe3D value-added catalog of spatially resolved and
integrated properties of galaxies for DR17 (Sanchez et al 2022; Lacerda et
al 2022).  We use the integrated stellar mass derived from fitting the MILES
stellar population synthesis model outputs (Vazdekis et al 2010) to spectra
from each spaxel to obtain the local stellar surface density $\Sigma_*$
and then integrating the local stellar surface density estimates to obtain
the total stellar mass. The total star formation rates are estimated in a
similar way using the H$\alpha$ emission line luminosities, corrected for
dust extinction using the value of the Balmer decrement $H\alpha/H\beta$. The
stellar mass estimates from the  Pipe3D catalogue tend to be systematically
higher by 0.1-0.2 dex because the contribution from lower surface brightness
regions of the galaxy are included.
Error bars on the slope and R$_{break}$ parameters are
calculated by adding noise to the profiles according to the flux errors
calculated by boot-strapping the spaxels considered in each radial bin. The
error bars on the slope are smaller than the size of the plotted symbols
and cannot be seen.  Some of the errors on R$_{_break}$ are more substantial
and can  be seen in the figure.

\begin{figure*}
\includegraphics[width=135mm]{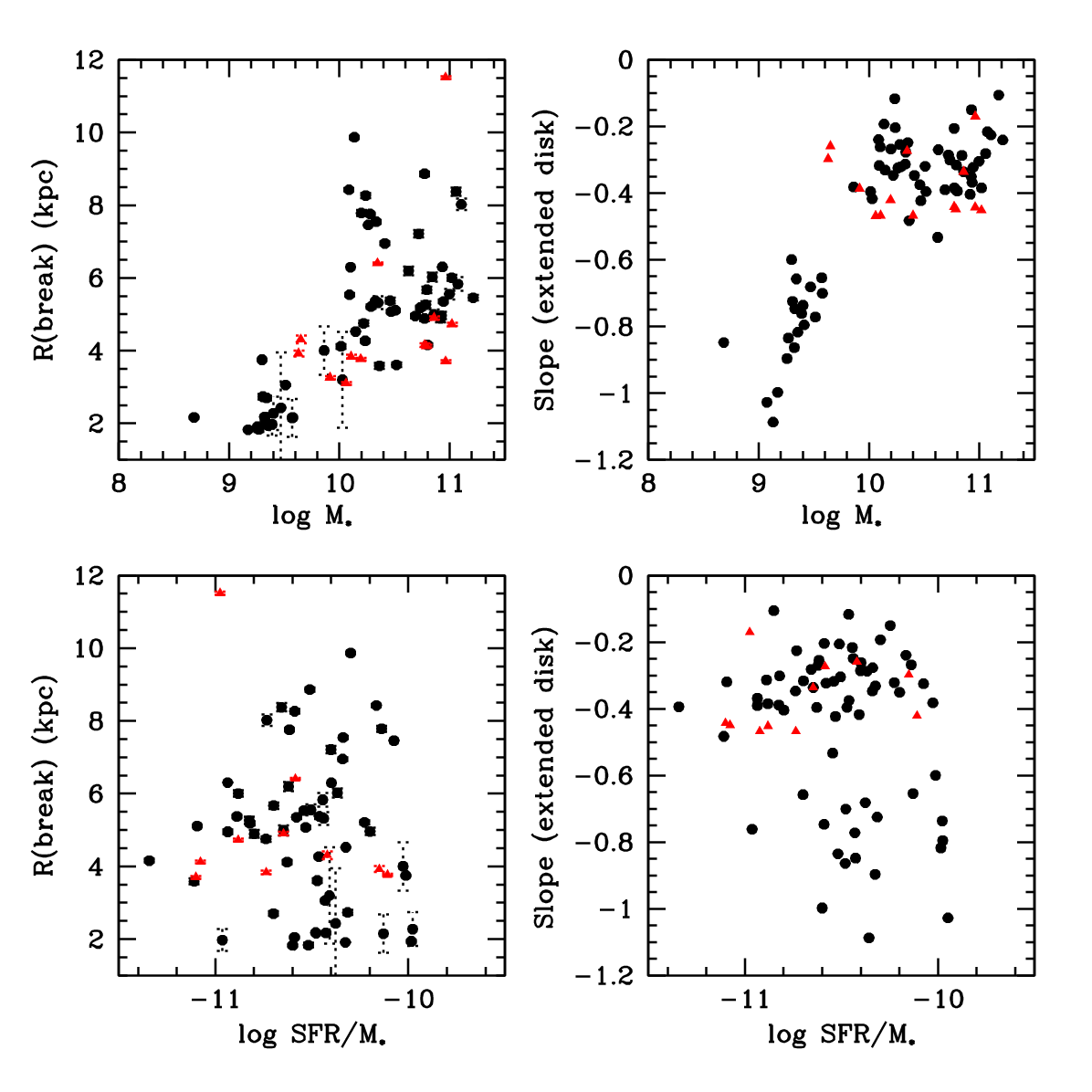}
\caption{ The two thick disk  profile parameters slope $b$ and
break radius, R$_{break}$, are plotted as
a function of  the stellar
mass and the logarithm of the  specific star formation rate $\log SFR/M_*$
for 82 galaxies in our sample. The disk-dominated galaxies are plotted as
black circles, while the sample with higher Sersic indices are plotted as
red triangles.
\label{models}}
\end{figure*}

Figure 7 shows that there is a  tight scaling between R$_{break}$ and
slope $b$ for galaxies in the stellar mass range $10^9 -10^{10} M_{\odot}$.
The Pearson correlation coefficient $R$ has a value of 0.73 with an associated
p-value of $3.85 \times 10^{-5}$. 
Above this stellar mass, there is no longer a strong correlation between
profile  parameters and stellar mass ($R$=0.11 and $p$=0.37), and the scatter at fixed $M_*$ also
increases.  

The association of the $R_{break}$ with the transition from the
thick disk to the halo component is very clearly demonstrated in recent
work by Gilhuly et al (2022) analyzing 12 nearby edge-on galaxies from
the Dragonfly Edge-on Galaxies Survey (DEGS).  These observations reached
limiting $g$ and $r$-band surface brightnesses of 32 mag/arcsec$^2$,
allowing the stellar light to be probed out to distances of 20-60 kpc from
the disk plane. These authors constructed surface brightness profiles from
elliptical isophote fitting as well as   profiles as a function of distance along the
minor axis, as done in this paper. The minor axis profiles shown in Figure
9 of Gilhuly et al  show a steep inner power-law region and a clear flattening
that begins  at similar radii as found in this analysis.

As noted by Gilhuly et al (2020), the profile breaks seen in edge-on galaxies
can only be used as a very approximate indicator of a transition from the
star-forming disk  to a ``pure'' stellar halo consisting of stars accreted
during the hierarchical assembly of the surrounding dark matter halo.
Studies of the stellar mass surface density profiles of 17 Milky Way-mass
galaxies from the FIRE-2 suite of simulations by Sanderson et al. (2018) found
no threshold or inflection point that consistently indicated a transition to
an accretion-dominated regime.  This was directly attributed to the natural
variety among accretion histories and therefore the amount and distribution
of accreted stars with respect to stars that were formed ``in situ'' from
gas in the disk.  The large observed  scatter in R$_{break}$ as a function
of stellar mass of the galaxy  that we find in the Milky Way mass range is
quite consistent with these  simulation predictions. 

At stellar masses less than
$10^{10} M_{\odot}$, we find that R$_{break}$ increases   as a
function of $M_*$ and  there is a   significant correlation between these 
quantities ($R$=0.566, $p=3.95\times10^{-3}$).
This  could be an indication that the relative contribution  of
the accreted component is smaller for lower mass galaxies. Wang \& Kauffmann
(2008, see Figure 3) use semi-analytic models implemented within the Millennium
simulation to show that fraction of galaxies that have not experienced either
major or minor mergers increases strongly at stellar masses less than $10^{10}
M_{\odot}$.  

Finally, the lack of correlation between the profile parameters
and the star formation activity in the disk, as indicated by $\log SFR/M_*$
is also an indication that these parameters are not determined by present-day
gas-physical processes in the disk.

The majority of the  subsample of galaxies with higher Sersic indices,
indicated by the filled red triangles in Figure 7  have stellar masses
greater than $10^{10} M_{\odot}$ and have profile parameters that overlap
those of the disk-dominated edge-on galaxies.  Only two galaxies in the
higher Sersic index sample have stellar masses $\sim 10^{9.5} M_{sol}$ and
interestingly, these have extended disk profiles with much shallower slopes
than  more disky edge-on galaxies of the same mass.  This may indicate that
the same process that created the bulge was responsible for creating a more
extended thick disk with a shallower light profile. Larger samples of low mass
edge-on galaxies are needed to confirm this conjecture.

\subsection {4000 \AA\ break profiles}

Our analysis of the 4000 \AA\ break profiles begins with a linear fit to the
D$_n$(4000) measurements within a distance of 1.5 kpc from the disk plane. A
``break'' in the profile is defined as the first time one of two conditions are
satisfied: 1) Four  successive  D$_n$(4000) measurements lie more than 3$\sigma$
away from the inner disk fit, 2) One  measurement lies more than  3$\sigma$ away
from the inner disk fit, and the next three D$_n$(4000) measurements exhibit
increasing deviation from the fit. The first condition identifies breaks in
regions of the profile with high $S/N$ D$_n$(4000) measurements, while the
second criterion is designed for lower $S/N$ outer region measurements.  Once a
break is identified, a new linear fit is carried out starting with  measurments 
more than 0.6 kpc from the breakpoint, and a new break-finding procedure is restarted. 
In Figure 8, we show an example of a galaxy where three breaks are identified in the
D$_n$(4000) profiles, with the position of each beak indicated by red vertical lines
on the plot.

\begin{figure}
\includegraphics[width=75mm]{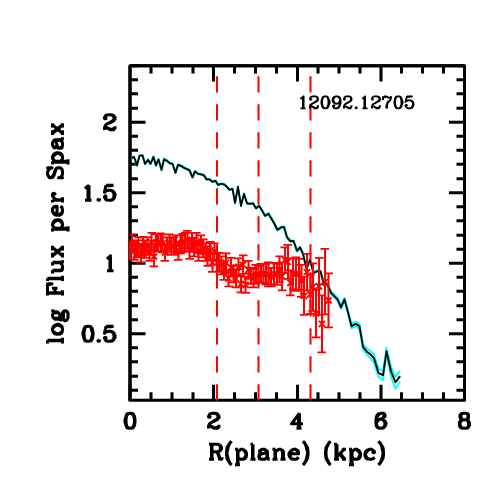}
An example of a galaxy where three breaks are identified in the
D$_n$(4000) profiles, with the position of each beak indicated by red vertical lines
on the plot.
\label{models}
\end{figure}

We define D$_n$(4000)(inner) as the average D$_n$(4000) value within the inner 1.5
kpc region of the galaxy, and D$_n$(4000)(outer) as the average D$_n$(4000)
value in the region starting from the last break point to 0.6 kpc beyond the
break point. The quantity  D$_n$(4000)(outer)-D$_n$(4000)(inner) provides a
measure of the gradient in age of the stellar population. We also record the
quantity $N_{break}$, the number of break points in the D$_n$(4000) profile,
as a measure of the irregularity in the stellar population distributions in
the outer disk and halo transition regions.

The two upper panels of Figure 9 show the relation between D$_n$(4000)(inner)
and the stellar mass and specific star formation rate of the host galaxies.
There is a strong correlation with SFR/$M_*$ ($R$=-0.72, p=$10^{-14}$),
and a weaker one with $M_*$ ($R$=0.50, p=$ 1.5 \times 10^{-6}$), reflecting the fact
that less massive disk-dominated galaxies have younger stellar populations.
For edge-on galaxies, dust extinction effects may also play a secondary role
in increasing the measured value of D$_n$(4000) at fixed stellar age, but we
note that we use the narrow version of the index that is measured within 100
\AA\ wide windows, so the effect of dust is minimized.  Edge-on galaxies with
higher Sersic indices (plotted as red triangle) exhibit somewhat more scatter
in the top panels of Figure 9, but otherwise follow the  same correlation
defined by the more disky edge-on galaxies.

In the bottom left panel, we plot  D$_n$(4000)(outer)-D$_n$(4000)(inner) as
a function of the stellar mass of the galaxy. The numbers in the left-hand
part of the figure indicate (from top to bottom) the number of galaxies where this quantity is
positive, zero (indicating a flat profile with no breaks) and negative. The
numbers in parentheses are for the high Sersic index sample. The results
indicate that the 4000 \AA\ break profiles are flat or positive for  85\%
of the sample, indicating that for the majority of galaxies,  the extended
disk regions contain stars that are as old or older than those in the inner
disk. These results are the same for the pure disk and  for the high Sersic
index samples.  The quantity that differs significantly between the disky
and the bulgey samples  is the distribution of N$_{break}$, the number
of break points in the D$_n$(4000) profile (plotted in the bottom right
panel of Figure 9). 40\% of the pure disk edge-on
galaxies have no breaks in their D$_n$(4000) profiles, whereas only 1 out
of 14 galaxies in the high Sersic index sample has no breaks. More than a
third of the high Sersic index sample has 2 or more breaks, whereas only 12\%
of the disky sample have this many breaks.  As we will discuss
in more detail later, the fact that the number of  breaks  correlates with
the morphological type of the galaxy and also the fact that they occur
most often at large extra-planar distances, may argue for a merger/accretion 
origin.  

\begin{figure*}
\includegraphics[width=135mm]{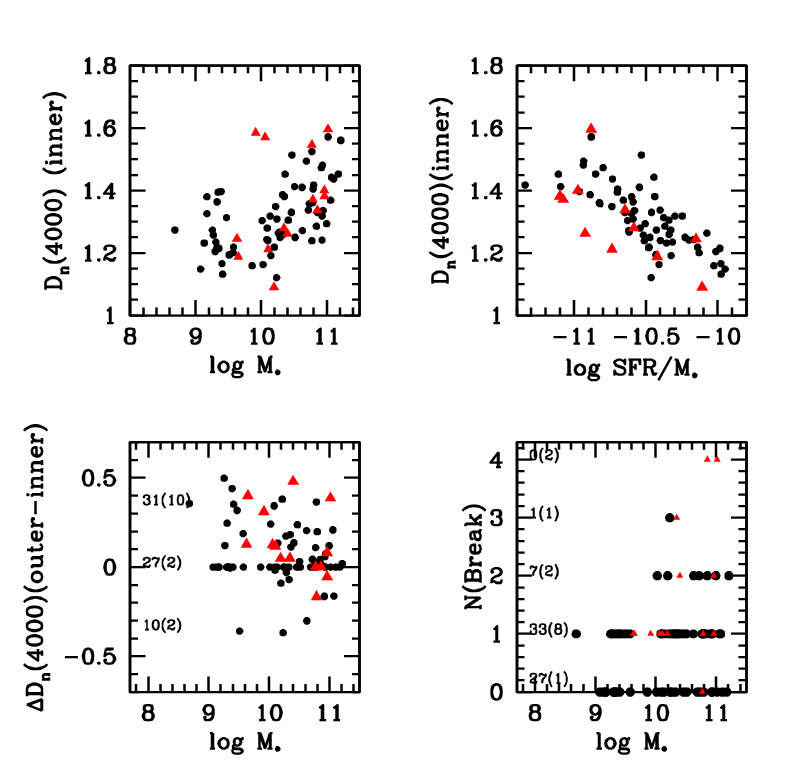}
\caption{The two upper panels show the relation between D$_n$(4000)(inner)
and the stellar mass and specific star formation rate of the host galaxies.
The bottom left panel shows  D$_n$(4000)(outer)-D$_n$(4000)(inner) as
a function of the stellar mass of the galaxy. The numbers in the left-hand
part of the figure (from top to bottom) indicate the number of galaxies where this quantity is
positive, zero (indicating a flat profile with no breaks) and negative. The
numbers in parentheses are for the high Sersic index sample. 
The bottom right panel shows N(break) as a function of the stellar mass.
 The numbers in the left-hand
part of the figure indicate the number of galaxies where this quantity
is 0,1,2,3 and 4, with the number in parentheses for the high
Sersic index sample.
\label{models}}
\end{figure*}

We now split the sample into galaxies with positive/zero values of
 D$_n$(4000)(outer)-D$_n$(4000)(inner) and negative value of this quantity.
Histograms of the stellar mass, the specific star formation rate of the
galaxies, as well as the slope $b$ of the power-law component of the extended
disk  are plotted in the top row of Figure 10 . The black histograms are for the
galaxies with  flat D$_n$(4000) profiles and the blue histograms are for the
galaxies where the stellar populations become younger in the outer disk. As
can be seen, there is no statistically significant difference between
the two subsamples.  In the bottom panels the sample is split according
to N$_{break}$, with black histograms for galaxies with no breaks and red
histograms for galaxies with one or more breaks. Here it can be seen that
galaxies with breaks have higher stellar masses
and shallower outer disk slopes.

\begin{figure*}
\includegraphics[width=160mm]{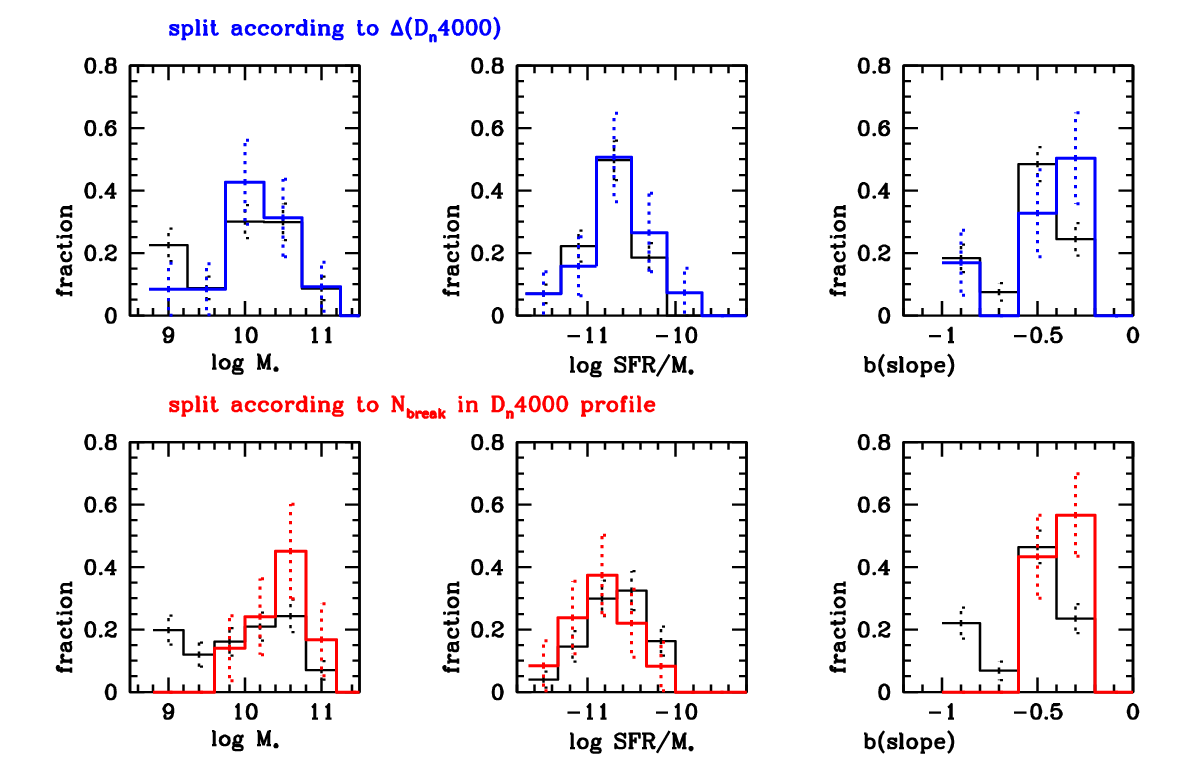}
\caption{ 
Histograms of the stellar mass, the specific star formation rate of the
galaxies, and the  slope $b$ of the power-law component of the extended
disk are plotted in the top row. The black histograms are for the
galaxies with  flat D$_n$(4000) profiles and the blue histograms are for the
galaxies where the stellar populations become younger in the outer disk
In the bottom row,  black histograms for galaxies with no breaks and red
histograms for galaxies with one or more breaks. Dotten lines show the error bars
calculated by bootstrap resampling.
\label{models}}
\end{figure*}

\subsection {Emission line properties in the outer disk and halo transition
regions}

Figures 11  and 12 show the equivalent widths of the [OII] and H$\alpha$
emission lines plotted as a function of stellar mass (Figure 11) and $\log
SFR/M_*$ (Figure 12).  The results are plotted for the inner disk stacked
spectra (left column), the extended disk stacked spectra (middle column)
and the halo transition region stacks (right column). There is no strong
correlation between  [OII] and H$\alpha$ equivalent widths and stellar mass
in any of the three regions. There is a weak tendency for the most massive
galaxies to have slightly weaker emission lines.  
Figure 12 shows that the emission line equivalent widths in all three regions
correlate with the specific star formation rate of the galaxy. The correlation in 
the halo transition region is somewhat clearer for the H$\alpha$ line than for the
[OII] line. 

\begin{figure*}
\includegraphics[width=160mm]{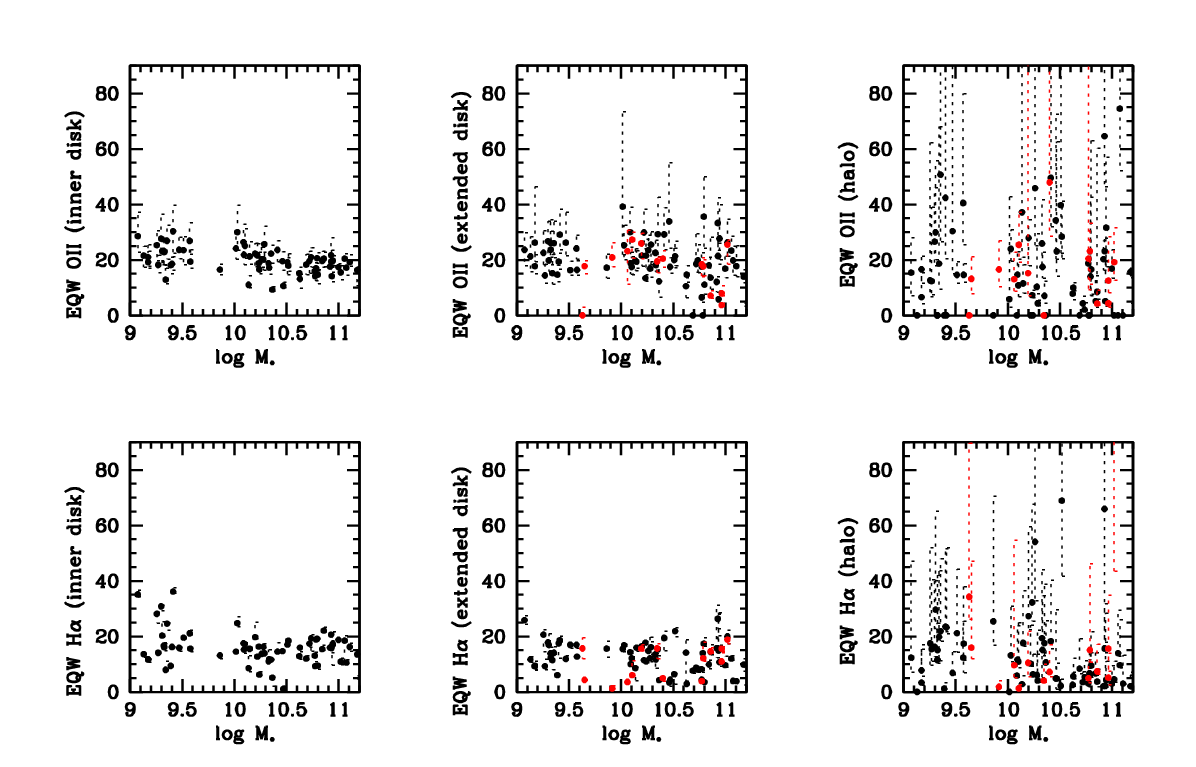}
\caption{ 
The equivalent widths of the [OII] and H$\alpha$
emission lines is plotted as a function of stellar mass. 
  The results are plotted for the inner disk stacked
spectra (left column), the extended disk stacked spectra (middle column)
and the halo transition region stacks (right column).
\label{models}}
\end{figure*}

\begin{figure*}
\includegraphics[width=160mm]{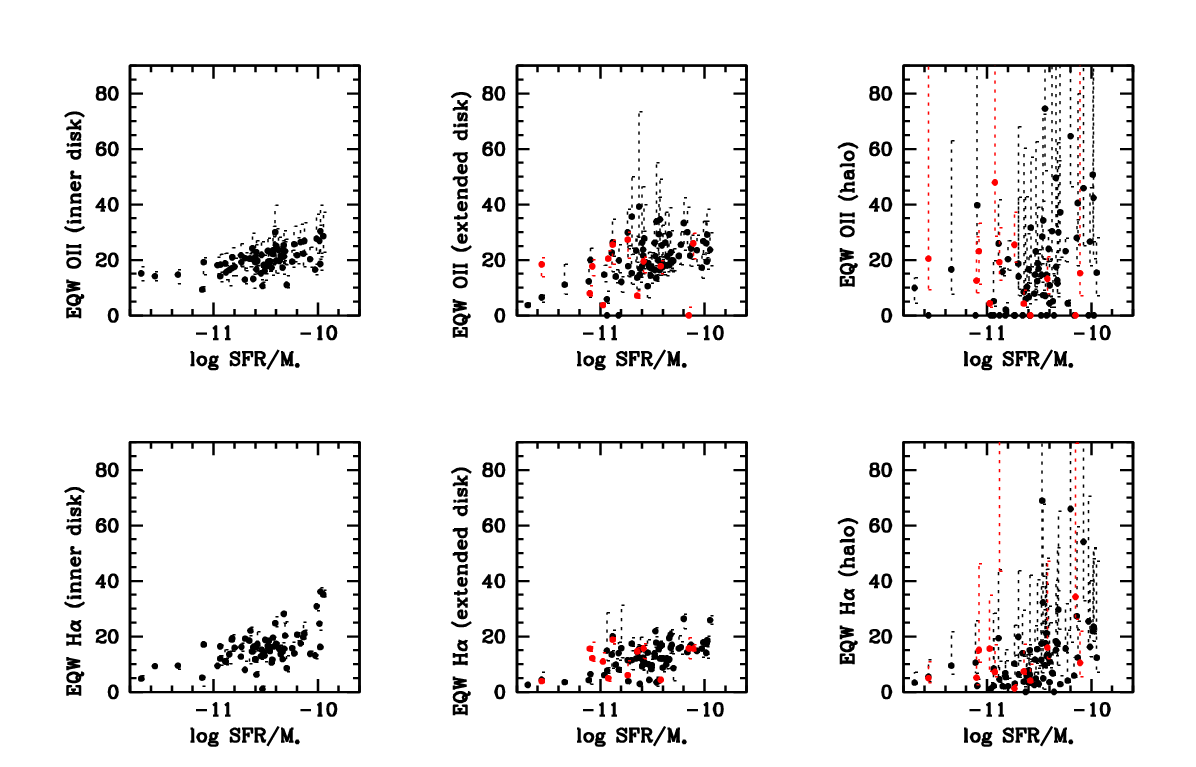}
\caption{ 
As in the previous figure, except the emission lines are
plotted as function of $\log SFR/M_*$
\label{models}}
\end{figure*}

Similar to the analysis in the previous section, we split the sample into 
galaxies with positive/negative values of EQW(H$\alpha$)(outer disk)-EQW(H$\alpha$)(inner disk)
and  EQW([OII](outer disk)- EQW[OII](inner disk) and plot histograms of galaxy properties. 
The results are shown in Figure 13. In the previous section, we found that the
distribution of stellar masses and specific star formation rates was the same for galaxies
split according to 4000 \AA\ break gradient. Here we find that more massive
galaxies and galaxies with low values of SFR/$M_*$ are more likely to have
lower H$\alpha$ EQW in the outer disk compared to the inner disk. 
The same trend is not found for EQW [OII]. 

\begin{figure*}
\includegraphics[width=135mm]{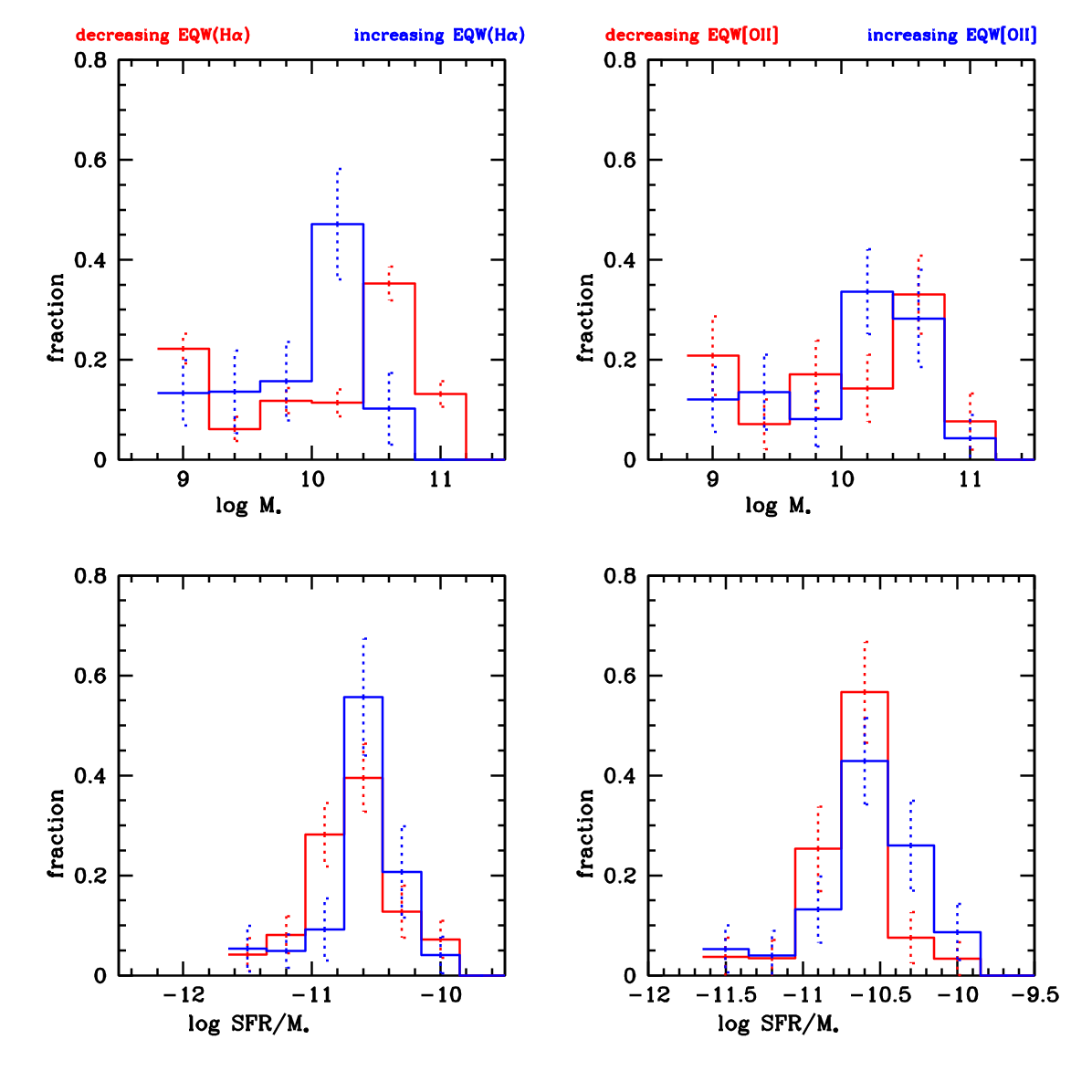}
\caption{ 
Histograms of galaxy properties for the sample
split according   positive/negative values of EQW(H$\alpha$)(outer)- EQW(H$\alpha$)(inner disk)
and  EQW([OII](outer)- EQW[OII](inner disk). 
\label{models}}
\end{figure*}

In Figure 14, we explore how the 
the [OII]/H$\alpha$ ratio changes  from the inner to the outer disk in more detail.  
In the  top left panel in the figure , we plot [OII]/H$\alpha$ in the outer 
disk as a function of [OII]/H$\alpha$ in the inner disk. The dashed line indicates the
one-to-one linear relation. Outer disk values are frequently, but not
always larger than the inner disk values.  In the next two panels, we plot
the ratio of outer-to-inner  [OII]/H$\alpha$ values as a function of
the stellar mass and the specific star formation rate of the galaxy. The results
in these two panels confirm that a large increase in [OII]/H$\alpha$ in the outer disk
is found in massive galaxies with low specific star formation rate. It is thus unlikely
that shocks due to supernovae-driven outflows or infalling cold gas is 
responsible for this pattern of ionization. In the bottom panels, we test whether 
the ratio of outer-to-inner disk [OII]/H$\alpha$ values  depends on the
properties of the outer disk. The bottom left panel shows that large increases  
in  [OII]/H$\alpha$ ratio occur in galaxies with extended thick disks (larger values of
the slope parameter  $b$).
A shallow slope is indicative of a higher fraction  of stars located 
external to the thin disk. The next panel shows that if the 4000 \AA\ break
strength decreases in the outer disk, the outer [OII]/H$\alpha$ ratios
also decrease. Conversely, most galaxies with increasing  4000 \AA\ break
values have increasing values of [OII]/H$\alpha$. 

In summary, the observational 
data  support  the hypothesis that ionizing radiation
from evolved stars is responsible for the increasing  [OII]/H$\alpha$ values
away from the mid-plane of the galactic disk in massive galaxies.   

\begin{figure*}
\includegraphics[width=160mm]{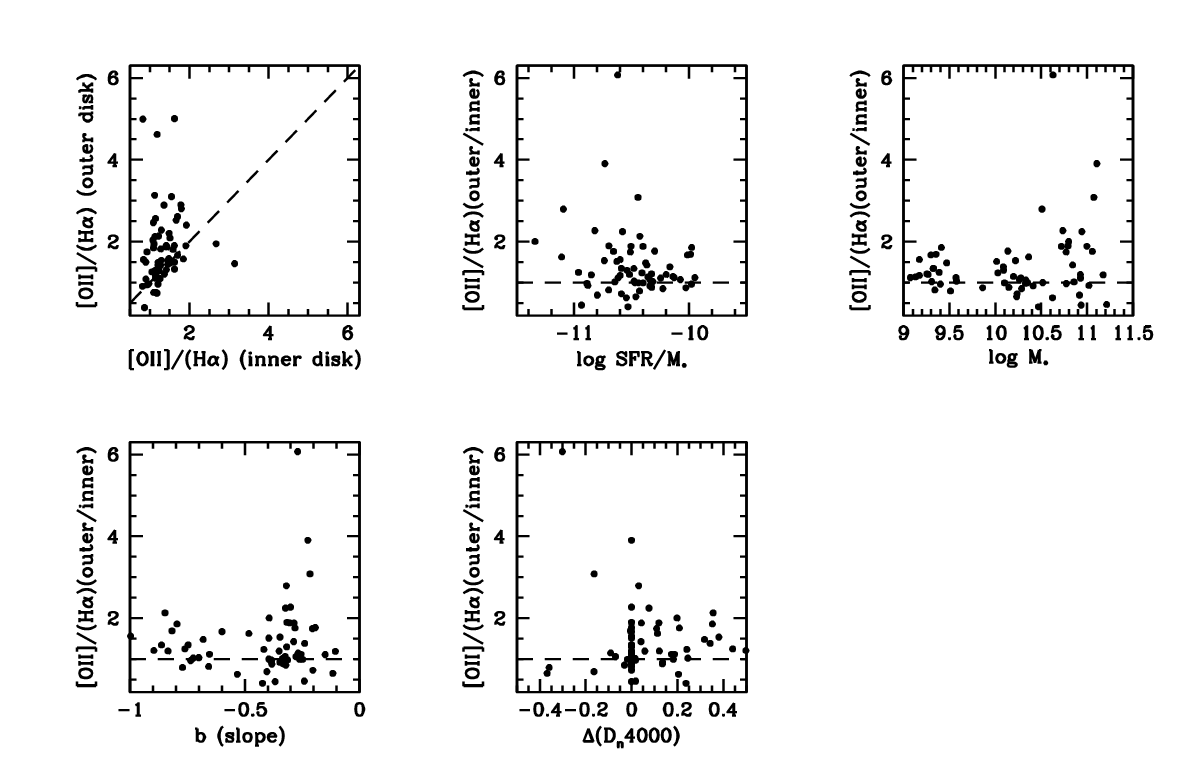}
\caption{[OII]/H$\alpha$ in the outer 
disk is plotted as a function of [OII]/H$\alpha$ in the inner disk. The dashed line indicates the
one-to-one linear relation.  The next two panels show
the ratio of outer disk to inner disk [OII]/H$\alpha$ values as a function of
the stellar mass and the specific star formation rate of the galaxy. 
The bottom  panels show
the ratio of outer disk to inner disk [OII]/H$\alpha$ values as a function of
the outer disk slope $b$) and as a function of 4000\AA\ break gradient.  
\label{models}}
\end{figure*}

\section {Summary and Discussion}

We have studied the stellar light, 4000 \AA\ break and emission line profiles of 
a sample of 82 edge-on galaxies from the MaNGA  survey.
We characterize the stellar light profiles perpendicular to the disk plane
using two parameters: a) the power-law slope of the  thick disk component, 2) the
break radius where the profile flattens. We do not attempt to
deconvolve the thin and thick disk components of the galaxy because of the
limited spatial resolution of the MaNGA IFU spaxel sampling. 

We characterize the 4000 \AA\ break profiles perpendicular to the disk plane
by the number of breaks (defined as significant changes in slope) in the profile and by the change in D$_n$(4000)
from the inner 1.5 kpc region of the  disk to the region just beyond the last break.
The emission line characteristics of the extraplanar gas are characterized  using measurements
of H$\alpha$ and [OII] EQW from stacked spectra  in the thin disk, the power-law thick disk and the halo
transition regions.

The main results of our analysis are the following:
\begin{itemize}
\item Both the slope and the break radius exhibit a tight correlation with the
stellar mass of the galaxy over the stellar mass range 
$10^9 < \log M_* < 10^{10} M_{\odot}$ At larger stellar masses,
both parameters no longer correlate with $M_*$ and also exhibit substantial
scatter at fixed $M_*$.
\item  50\% of the sample have thick disks with larger values of D$_n$(4000) than  the inner disk.
The change in break strength is uniformly spread  over values ranging from close to zero to 0.5.
35\% of the sample have flat D$_n$(4000) profiles and  15\% of the sample have 
smaller D$_n$(4000) values in thick disk. The shift towards lower break values is generally small
in these systems. 
Very ``young'' thick disk components are thus rare. 
\item The D$_n$(4000) profiles exhibit up to 4 separate breaks. The number of breaks
is higher in galaxies with larger stellar masses, in galaxies with more extended thick
disks and in galaxies with a  larger bulge component. We hypothesize that the
breaks are produced by accretion events, which have been  more frequent in massive
galaxies with bulges.
\item In contrast, the change in 4000 \AA\ break strength from the inner disk to the outer
regions of the thick disk does not correlate with any of the global or structural properties
of the galaxy. This may suggest that the accreted satellites that build the thick disk span a wide
range in age and/or  metallicity. 
\item  
In agreement with previous studies, the extraplanar H$\alpha$ EQW  correlates most strongly
with the specific star formation rate of the galaxy, and the [OII]/H$\alpha$ ratio  increases
with  distance from the disk plane. This increase is most apparent for the most
massive galaxies with the lowest SFR/$M_*$ and the most extended stellar thick disk 
components. These findings support the hypothesis that the larger [OII]/H$\alpha$ ratios
may be caused by ionization from evolved stars, rather than by other mechanisms
such as shocks.
\end{itemize}

In Figure 15, we show tracks predicted by stellar population synthesis models 
in the plane of stellar mass-weighted age  versus D$-n$(4000).
The blue, black and red curves are for models with 0.25 solar, half solar and solar metallicity.
Since the aim is to interpret 4000 \AA\ break measurements for stacked spectra over
large spatial scales (7-20 kpc in diameter), we generate models with a continuous
star formation history beginning at  redshift $z=5$, with variable e-folding timescale
ranging from 0.1-13 Gyr. The population synthesis models are from Bruzual \& Charlot (2003)
assuming a Kroupa (2001) initial mass function.

We have marked the typical inner disk 4000 \AA\  break strength
of 1.4 (see Figure 4) as a solid line. T
The typical change in D$_n$(4000) over a break
region is around 0.1 (marked as dotted lines on the plot). It can be seen that
this corresponds to a shift in mean stellar age of around 1 Gyr. The sensitivity of 
D$_n$(4000) to metallicity is quite weak at younger stellar ages. Keeping the
stellar age fixed and shifting the metallicity by a factor of two only produces a
few percent rather than a ten percent change in the D$_n$(4000) index. The very largest
changes in D$_n$(4000) from 1.4 in the inner disk to 1.9 in the outermost regions where
the index is still measurable, can be explained by an increase in mean stellar age of 
around 5 Gyr.

\begin{figure}
\includegraphics[width=90mm]{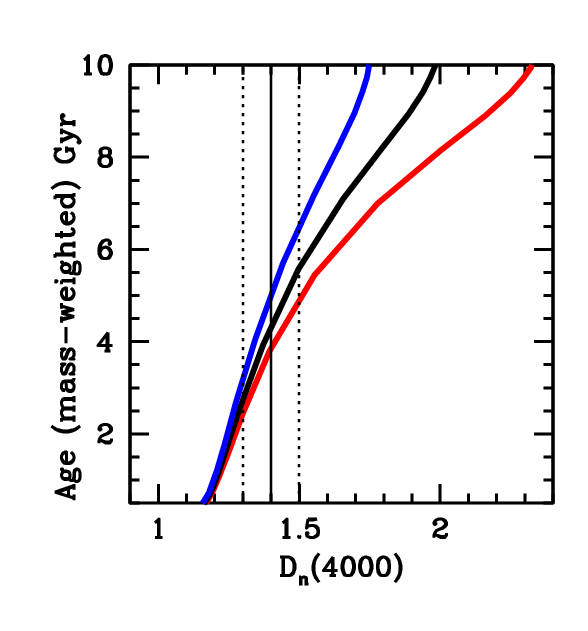}
\caption{
Tracks in the plane of stellar mass-weighted age  versus D$-n$(4000).
The blue, black and red curves are for models with 0.25 solar, half solar and solar metallicity.
The typical inner disk 4000 \AA\ break strength                            
of 1.4 is marked  as a solid line. The typical change in D$_n$(4000) over a break
region of  0.1 is marked as dotted lines on the plot). 
\label{models}}
\end{figure}
 
These findings are in good agreement with the results of 
Sattler et al (2024), who find an average age difference of 2 Gyr between the thin
and thick disks of a sample of eight edge-on star-forming disc galaxies observed with
the Multi-Unit Spectroscopic Explorer (MUSE). We note that the S/N that we are able to achieve with
our stacked MaNGA spectra does not allow us to study metallicity and [Mg/Fe] trends as
done in this paper.

There are two main physical mechanisms that have been demonstrated  
to produce thick disks in cosmological simulations of galaxy formation. In
N-body simulations that do not include gas, the impact of satellite accretion
events that can create thick disks can be  assessed by studying
the frequency of dark matter subhalo merging events that are aligned with
the disk plane. In such studies,  the disk plane is assumed to be set by the angular momentum
vector of the main halo (e.g. Read et al 2008).

The modern generation of cosmological hydrodynamical simulations are
much more successful in producing realistic Milky-Way type galaxies and
it is generally agreed that a combination of gas-physical processes and
satellite accretion events can contribute to the build-up of thick disks,
with gas-physical processes more important at earlier epochs.

Most of the published work has concentrated on Milky Way mass galaxies. Pinna
et al (2024) studied thick disk formation in the AURIGA simulations of 24
Milky Way type galaxies to assess the role played by satellite mergers. They
found that mergers contribute an average of 22\% of the stellar mass in thick
disks. In two of the galaxies, half of the thick disk mass was from accreted
material. In the majority of galaxies in the simulation, the thick disks were
found to be older, more metal poor and more enhanced in alpha-elements with
respect to iron, indicating that the in situ stars were formed at earlier
epochs in the thick disk compared to the thin disk.

Other studies have concentrated on the detailed formation mechanisms of the
"in situ" component. Yu et al (2021) emphasized the role of early bursts of
star formation in driving thick disk formation in Milky Way type galaxies
in the FIRE-2 cosmological simulations. Most recently Chandra et al (2024)
put forward a three-phase evolutionary scenario for our own  Milky Way:
the disordered and chaotic protogalaxy that forms the bulge, followed by
a kinematically hot old disk, and finally a cool-down phase that forms the
kinematically cold young disk.

Our work confirms the standard picture for Milky Way type galaxies of a
thick disk component that is older than the thin disk. None of the structural
or stellar population  properties  of the thick disk are found to correlate
with the present day star formation activity in the thin disk, indicating
that the bulk of the thick disk formed at earlier epochs. There is also
substantial scatter in the relation between the slope and   break radius of the thick disk and
the stellar mass of the galaxy over the range
$10<\log M_* < 11$. This is likely indicative of significant
variation in satellite accretion histories between different objects, also
in line with simulation predictions. Interestingly,
our results also ech the  diversity found in stellar halo properties of galaxies
in the same stellar mass range (Harmsen et al. 2017,  Monachesi et al. 2019, Gillhuly et al. 2022).  

Our results demonstrate  that the structural properties
of lower mass disk galaxies differ substantially from those of Milky Way
mass galaxies. 
The lower the mass of the galaxy, the steeper the surface density
profile of its thick disk, indicating a smaller contribution to the total
disk stellar mass. Since it has been demonstrated that the probability of
a bursty star formation history is {\em higher} for low mass galaxies (see
for example Kauffmann 2014), this
calls into question the hypothesis that starbursts are responsible for thick dick formation
in low mass  systems. On the other hand, the number of merger/accretion
events is smaller for low mass galaxies, so the observed trends may me mainly 
driven by increasing quiescent accretion histories. High resolution simulations
of low mass galaxies will be a valuable tool in testing whether the low mass
galaxy scaling relation are easily explainable.

Finally, we note that it is important to understand whether  
the breaks in the D$_n$(4000) profiles are a fossil record of past merging
events. 
Once again, N-body+hydrodynamical simulation studies with
mock IFU cubes smoothed to the same resolution as the 
MaNGA cubes  would be valuable way to try and understand the origin of such breaks.
Such studies would also allow us to assess whether  the next generation of deep imaging surveys 
with the EUCLID telescope in space and the 
Vera C. Rubin Observatory on the ground would be able to localize the source of the
D$_n$(4000) discontinuities, delineate the morphology of the
regions where stellar population properties change abruptly, 
and help us understand the detailed accretion
history of galaxies and their halos in more detail.

\vspace{4mm}
{\bf Acknowledgements}\\
\normalsize

G.K.  thanks Claude-Andr\'e Faucher-Gigu\'ere 
and Jing Wang for useful discussions.
A.M. gratefully acknowledges support by the 
ANID BASAL project FB210003, by the FONDECYT Regular 
grant 1212046, and funding from the Max Planck Society 
through a "PartnerGroup" grant.
Funding for SDSS-IV has been provided by the Alfred
P. Sloan Foundation and Participating Institutions. Ad-
ditional funding towards SDSS-IV has been provided by
the US Department of Energy Onece of Science. SDSS-
IV acknowledges support and resources from the Centre
for High-Performance Computing at the University of
Utah. The SDSS web site is www.sdss.org.
SDSS-IV is managed by the Astrophysical Research
Consortium for the Participating Institutions of the
SDSS Collaboration including the Brazilian Participation
Group, the Carnegie Institution for Science,
Carnegie Mellon University, the Chilean Participation
Group, the French Participation Group, Harvard-
Smithsonian Center for Astrophysics, Instituto de
Astrofsica de Canarias, The Johns Hopkins University
sity, Kavli Institute for the Physics and Mathematics
of the Universe (IPMU)/University of Tokyo, Lawrence
Berkeley National Laboratory, Leibniz Institut fur
Astrophysik Potsdam (AIP), Max-Planck-Institut f\"ur
Astronomie (MPIA Heidelberg), Max-Planck-Institut f\"ur
Astrophysik (MPA Garching), Max-Planck-Institut f\"ur
Extraterrestrische Physik (MPE), National Astronom-
ical Observatory of China, New Mexico State University,
New York University, University of Notre Dame,
Observatario Nacional/MCTI, the Ohio State University,
Pennsylvania State University, Shanghai Astronomical
Observatory, United Kingdom Participation Group,
Universidad Nacional Autonoma de Mexico, University
of Arizona, University of Colorado Boulder, University
of Oxford, University of Portsmouth, University of Utah,
University of Virginia, University of Washington,
University of Wisconsin, Vanderbilt University and Yale
University.

\vspace{4mm}
{\bf Data Availability}\\
\normalsize
Data from this paper will be made available on reasonable request to the
corresponding author.

%===================================


\begin{thebibliography}{}

\bibitem[\protect\citeauthoryear{Abadi, Ben{\'\i}tez-Llambay, \& Ferrero}{2013}]{2013BAAA...56...33A} Abadi M.~G., Ben{\'\i}tez-Llambay A., Ferrero I., 2013, BAAA, 56, 33

\bibitem[\protect\citeauthoryear{Abdurro'uf et al.}{2022}]{2022ApJS..259...35A} Abdurro'uf, Accetta K., Aerts C., Silva Aguirre V., Ahumada R., Ajgaonkar N., Filiz Ak N., et al., 2022, ApJS, 259, 35. doi:10.3847/1538-4365/ac4414

\bibitem[\protect\citeauthoryear{Aguado et al.}{2019}]{2019ApJS..240...23A} Aguado D.~S., Ahumada R., Almeida A., Anderson S.~F., Andrews B.~H., Anguiano B., Aquino Ort{\'\i}z E., et al., 2019, ApJS, 240, 23. doi:10.3847/1538-4365/aaf651

\bibitem[\protect\citeauthoryear{Bekki \& Chiba}{2001}]{2001ApJ...558..666B} Bekki K., Chiba M., 2001, ApJ, 558, 666. doi:10.1086/322300

\bibitem[\protect\citeauthoryear{Beom et al.}{2022}]{2022MNRAS.516.3175B} Beom M., Bizyaev D., Walterbos R.~A.~M., Chen Y., 2022, MNRAS, 516, 3175. doi:10.1093/mnras/stac1499

\bibitem[\protect\citeauthoryear{Blanton et al.}{2017}]{2017AJ....154...28B} Blanton M.~R., Bershady M.~A., Abolfathi B., Albareti F.~D., Allende Prieto C., Almeida A., Alonso-Garc{\'\i}a J., et al., 2017, AJ, 154, 28. doi:10.3847/1538-3881/aa7567

\bibitem[\protect\citeauthoryear{Brook et al.}{2004}]{2004ApJ...612..894B} Brook C.~B., Kawata D., Gibson B.~K., Freeman K.~C., 2004, ApJ, 612, 894. doi:10.1086/422709

\bibitem[\protect\citeauthoryear{Brook et al.}{2012}]{2012MNRAS.426..690B} Brook C.~B., Stinson G.~S., Gibson B.~K., Kawata D., House E.~L., Miranda M.~S., Macci{\`o} A.~V., et al., 2012, MNRAS, 426, 690. doi:10.1111/j.1365-2966.2012.21738.x


\bibitem[\protect\citeauthoryear{Bruzual \& Charlot}{2003}]{2003MNRAS.344.1000B} Bruzual G., Charlot S., 2003, MNRAS, 344, 1000. doi:10.1046/j.1365-8711.2003.06897.x

\bibitem[\protect\citeauthoryear{Bundy et al.}{2015}]{2015ApJ...798....7B} Bundy K., Bershady M.~A., Law D.~R., Yan R., Drory N., MacDonald N., Wake D.~A., et al., 2015, ApJ, 798, 7. doi:10.1088/0004-637X/798/1/7

\bibitem[\protect\citeauthoryear{Chandra et al.}{2024}]{2024ApJ...972..112C} Chandra V., Semenov V.~A., Rix H.-W., Conroy C., Bonaca A., Naidu R.~P., Andrae R., et al., 2024, ApJ, 972, 112. doi:10.3847/1538-4357/ad5b60

\bibitem[\protect\citeauthoryear{Comer{\'o}n et al.}{2011}]{2011ApJ...741...28C} Comer{\'o}n S., Elmegreen B.~G., Knapen J.~H., Salo H., Laurikainen E., Laine J., Athanassoula E., et al., 2011, ApJ, 741, 28. doi:10.1088/0004-637X/741/1/28

\bibitem[\protect\citeauthoryear{Comer{\'o}n et al.}{2014}]{2014A&A...571A..58C} Comer{\'o}n S., Elmegreen B.~G., Salo H., Laurikainen E., Holwerda B.~W., Knapen J.~H., 2014, A\&A, 571, A58. doi:10.1051/0004-6361/201424412

\bibitem[\protect\citeauthoryear{Dalcanton \& Bernstein}{2000}]{2000AJ....120..203D} Dalcanton J.~J., Bernstein R.~A., 2000, AJ, 120, 203. doi:10.1086/301425

\bibitem[\protect\citeauthoryear{Drory et al.}{2015}]{2015AJ....149...77D} Drory N., MacDonald N., Bershady M.~A., Bundy K., Gunn J., Law D.~R., Smith M., et al., 2015, AJ, 149, 77. doi:10.1088/0004-6256/149/2/77

\bibitem[\protect\citeauthoryear{Elias et al.}{2018}]{2018MNRAS.479.4004E} Elias L.~M., Sales L.~V., Creasey P., Cooper M.~C., Bullock J.~S., Rich R.~M., Hernquist L., 2018, MNRAS, 479, 4004. doi:10.1093/mnras/sty1718


\bibitem[\protect\citeauthoryear{Freeman}{1974}]{1974IAUS...58..129F} Freeman K.~C., 1974, IAUS, 58, 129

\bibitem[\protect\citeauthoryear{Gilhuly et al.}{2022}]{2022ApJ...932...44G} Gilhuly C., Merritt A., Abraham R., Danieli S., Lokhorst D., Liu Q., van Dokkum P., et al., 2022, ApJ, 932, 44. doi:10.3847/1538-4357/ac6750

\bibitem[\protect\citeauthoryear{Gilmore \& Reid}{1983}]{1983MNRAS.202.1025G} Gilmore G., Reid N., 1983, MNRAS, 202, 1025. doi:10.1093/mnras/202.4.1025

\bibitem[\protect\citeauthoryear{Gunn}{1982}]{1982ac...proc..133G} Gunn J.~E., 1982, ac...proc, 133

\bibitem[\protect\citeauthoryear{Gunn et al.}{2006}]{2006AJ....131.2332G} Gunn J.~E., Siegmund W.~A., Mannery E.~J., Owen R.~E., Hull C.~L., Leger R.~F., Carey L.~N., et al., 2006, AJ, 131, 2332. doi:10.1086/500975

\bibitem[\protect\citeauthoryear{Harmsen et al.}{2017}]{2017MNRAS.466.1491H} Harmsen B., Monachesi A., Bell E.~F., de Jong R.~S., Bailin J., Radburn-Smith D.~J., Holwerda B.~W., 2017, MNRAS, 466, 1491. doi:10.1093/mnras/stw2992

\bibitem[\protect\citeauthoryear{Ho et al.}{2016}]{2016MNRAS.457.1257H} Ho I.-T., Medling A.~M., Bland-Hawthorn J., Groves B., Kewley L.~J., Kobayashi C., Dopita M.~A., et al., 2016, MNRAS, 457, 1257. doi:10.1093/mnras/stw017

\bibitem[\protect\citeauthoryear{Jones et al.}{2017}]{2017A&A...599A.141J} Jones A., Kauffmann G., D'Souza R., Bizyaev D., Law D., Haffner L., Bah{\'e} Y., et al., 2017, A\&A, 599, A141. doi:10.1051/0004-6361/20162980

\bibitem[\protect\citeauthoryear{Karachentsev, Karachentseva, \& Parnovskij}{1993}]{1993AN....314...97K} Karachentsev I.~D., Karachentseva V.~E., Parnovskij S.~L., 1993, AN, 314, 97. doi:10.1002/asna.2113140302

\bibitem[\protect\citeauthoryear{Kauffmann et al.}{2003}]{2003MNRAS.341...54K} Kauffmann G., Heckman T.~M., White S.~D.~M., Charlot S., Tremonti C., Peng E.~W., Seibert M., et al., 2003, MNRAS, 341, 54. doi:10.1046/j.1365-8711.2003.06292.x

\bibitem[\protect\citeauthoryear{Kauffmann}{2014}]{2014MNRAS.441.2717K} Kauffmann G., 2014, MNRAS, 441, 2717. doi:10.1093/mnras/stu752

\bibitem[\protect\citeauthoryear{Kroupa}{2001}]{2001MNRAS.322..231K} Kroupa P., 2001, MNRAS, 322, 231. doi:10.1046/j.1365-8711.2001.04022.x

\bibitem[\protect\citeauthoryear{Lacerda et al.}{2022}]{2022NewA...9701895L} Lacerda E.~A.~D., S{\'a}nchez S.~F., Mej{\'\i}a-Narv{\'a}ez A., Camps-Fari{\~n}a A., Espinosa-Ponce C., Barrera-Ballesteros J.~K., Ibarra-Medel H., et al., 2022, NewA, 97, 101895. doi:10.1016/j.newast.2022.101895

\bibitem[\protect\citeauthoryear{Law et al.}{2015}]{2015AJ....150...19L} Law D.~R., Yan R., Bershady M.~A., Bundy K., Cherinka B., Drory N., MacDonald N., et al., 2015, AJ, 150, 19. doi:10.1088/0004-6256/150/1/19

\bibitem[\protect\citeauthoryear{Lehnert \& Heckman}{1996}]{1996ApJ...462..651L} Lehnert M.~D., Heckman T.~M., 1996, ApJ, 462, 651. doi:10.1086/177180

\bibitem[\protect\citeauthoryear{Merritt et al.}{2016}]{2016ApJ...830...62M} Merritt A., van Dokkum P., Abraham R., Zhang J., 2016, ApJ, 830, 62. doi:10.3847/0004-637X/830/2/62


\bibitem[\protect\citeauthoryear{Monachesi et al.}{2019}]{2019MNRAS.485.2589M} Monachesi A., G{\'o}mez F.~A., Grand R.~J.~J., Simpson C.~M., Kauffmann G., Bustamante S., Marinacci F., et al., 2019, MNRAS, 485, 2589. doi:10.1093/mnras/stz538

\bibitem[\protect\citeauthoryear{Pinna et al.}{2024}]{2024A&A...683A.236P} Pinna F., Walo-Mart{\'\i}n D., Grand R.~J.~J., Martig M., Fragkoudi F., G{\'o}mez F.~A., Marinacci F., et al., 2024, A\&A, 683, A236. doi:10.1051/0004-6361/202347388

\bibitem[\protect\citeauthoryear{Pohlen et al.}{2004}]{2004A&A...422..465P} Pohlen M., Balcells M., L{\"u}tticke R., Dettmar R.-J., 2004, A\&A, 422, 465. doi:10.1051/0004-6361:20035932

\bibitem[\protect\citeauthoryear{Quinn, Hernquist, \& Fullagar}{1993}]{1993ApJ...403...74Q} Quinn P.~J., Hernquist L., Fullagar D.~P., 1993, ApJ, 403, 74. doi:10.1086/172184

\bibitem[\protect\citeauthoryear{Read et al.}{2008}]{2008MNRAS.389.1041R} Read J.~I., Lake G., Agertz O., Debattista V.~P., 2008, MNRAS, 389, 1041. doi:10.1111/j.1365-2966.2008.13643.x

\bibitem[\protect\citeauthoryear{S{\'a}nchez et al.}{2022}]{2022ApJS..262...36S} S{\'a}nchez S.~F., Barrera-Ballesteros J.~K., Lacerda E., Mej{\'\i}a-Narvaez A., Camps-Fari{\~n}a A., Bruzual G., Espinosa-Ponce C., et al., 2022, ApJS, 262, 36. doi:10.3847/1538-4365/ac7b8f

\bibitem[\protect\citeauthoryear{Sanderson et al.}{2018}]{2018ApJ...869...12S} Sanderson R.~E., Garrison-Kimmel S., Wetzel A., Keung Chan T., Hopkins P.~F., Kere{\v{s}} D., Escala I., et al., 2018, ApJ, 869, 12. doi:10.3847/1538-4357/aaeb33

\bibitem[\protect\citeauthoryear{Sattler et al.}{2024}]{2024arXiv241005761S} Sattler N., Pinna F., Comer{\'o}n S., Martig M., Falc{\'o}n-Barroso J., Mart{\'\i}n-Navarro I., Neumayer N., 2024, arXiv, arXiv:2410.05761. doi:10.48550/arXiv.2410.05761


\bibitem[\protect\citeauthoryear{Smee et al.}{2013}]{2013AJ....146...32S} Smee S.~A., Gunn J.~E., Uomoto A., Roe N., Schlegel D., Rockosi C.~M., Carr M.~A., et al., 2013, AJ, 146, 32. doi:10.1088/0004-6256/146/2/32

\bibitem[\protect\citeauthoryear{Toth \& Ostriker}{1992}]{1992ApJ...389....5T} Toth G., Ostriker J.~P., 1992, ApJ, 389, 5. doi:10.1086/171185


\bibitem[\protect\citeauthoryear{van der Kruit}{1988}]{1988A&A...192..117V} van der Kruit P.~C., 1988, A\&A, 192, 117

\bibitem[\protect\citeauthoryear{van der Kruit \& Searle}{1981}]{1981A&A....95..105V} van der Kruit P.~C., Searle L., 1981, A\&A, 95, 105

\bibitem[\protect\citeauthoryear{Vazdekis et al.}{2010}]{2010MNRAS.404.1639V} Vazdekis A., S{\'a}nchez-Bl{\'a}zquez P., Falc{\'o}n-Barroso J., Cenarro A.~J., Beasley M.~A., Cardiel N., Gorgas J., et al., 2010, MNRAS, 404, 1639. doi:10.1111/j.1365-2966.2010.16407.x

\bibitem[\protect\citeauthoryear{Velazquez \& White}{1999}]{1999MNRAS.304..254V} Velazquez H., White S.~D.~M., 1999, MNRAS, 304, 254. doi:10.1046/j.1365-8711.1999.02354.x

\bibitem[\protect\citeauthoryear{Wake et al.}{2017}]{2017AJ....154...86W} Wake D.~A., Bundy K., Diamond-Stanic A.~M., Yan R., Blanton M.~R., Bershady M.~A., S{\'a}nchez-Gallego J.~R., et al., 2017, AJ, 154, 86. doi:10.3847/1538-3881/aa7ecc

\bibitem[\protect\citeauthoryear{Wang \& Kauffmann}{2008}]{2008MNRAS.391..785W} Wang L., Kauffmann G., 2008, MNRAS, 391, 785. doi:10.1111/j.1365-2966.2008.13907.x

\bibitem[\protect\citeauthoryear{Wu et al.}{2002}]{2002AJ....123.1364W} Wu H., Burstein D., Deng Z., Zhou X., Shang Z., Zheng Z., Chen J., et al., 2002, AJ, 123, 1364. doi:10.1086/338849

\bibitem[\protect\citeauthoryear{Yan et al.}{2016}]{2016AJ....152..197Y} Yan R., Bundy K., Law D.~R., Bershady M.~A., Andrews B., Cherinka B., Diamond-Stanic A.~M., et al., 2016, AJ, 152, 197. doi:10.3847/0004-6256/152/6/197

\bibitem[\protect\citeauthoryear{Yoachim \& Dalcanton}{2006}]{2006AJ....131..226Y} Yoachim P., Dalcanton J.~J., 2006, AJ, 131, 226. doi:10.1086/497970

\bibitem[\protect\citeauthoryear{Yu et al.}{2021}]{2021MNRAS.505..889Y} Yu S., Bullock J.~S., Klein C., Stern J., Wetzel A., Ma X., Moreno J., et al., 2021, MNRAS, 505, 889. doi:10.1093/mnras/stab1339



\end{thebibliography}
\end{document}